\documentclass[singlecolumn,epjc3]{svjour3}
\smartqed  
\RequirePackage{graphicx,mathtools,makecell,array,microtype,lineno,hyperref,hhline}
\usepackage{float}
\usepackage{color}
\usepackage[caption = false]{subfig}
\journalname{Eur. Phys. J. C}

\begin{document}

\title{Anisotropic strange stars in Tolman-Kuchowicz spacetime}

\author{M.K. Jasim\thanksref{e1,addr1}
\and Debabrata Deb\thanksref{e2,addr2}
\and Saibal Ray\thanksref{e3,addr3}
\and Y.K. Gupta\thanksref{e4,addr4}
\and Sourav Roy Chowdhury\thanksref{e5,addr2}.}

\thankstext{e1}{e-mail: mahmoodkhalid@unizwa.edu.om}
\thankstext{e2}{e-mail: ddeb.rs2016@physics.iiests.ac.in}
\thankstext{e3}{e-mail: saibal@associates.iucaa.in}
\thankstext{e4}{e-mail: kumar$001947$@gmail.com}
\thankstext{e5}{e-mail: sourav.rs2016@physics.iiests.ac.in}

\institute{Department of Mathematical and Physical Sciences, College of Arts and Science,
University of Nizwa, Nizwa, Sultanate of Oman\label{addr1}
\and Department of Physics, Indian Institute of Engineering Science and Technology, Shibpur, Howrah 711103, West Bengal, India\label{addr2}
\and Department of Physics, Government College of Engineering and Ceramic Technology, Kolkata 700010, West Bengal, India\label{addr3}
\and Department of Mathematics, Raj Kumar Goel Institute of Technology, Ghaziabad, 201003, Uttar Pradesh, India\label{addr4}}

\date{Received: date / Accepted: date}

\maketitle

\begin{abstract}
We attempt to study a singularity-free model for the spherically symmetric anisotropic strange stars under Einstein's general theory of relativity by exploiting the Tolman-Kuchowicz~\cite{Tolman1939,Kuchowicz1968} metric. Further, we have assumed that the cosmological constant $\Lambda$ is a scalar variable dependent on the spatial coordinate $r$. To describe the strange star candidates we have considered that they are made of strange quark matter (SQM) distribution, which is assumed to be governed by the MIT bag equation of state. To obtain unknown constants of the stellar system we match the interior Tolman-Kuchowicz metric to the exterior modified Schwarzschild metric with the cosmological constant, at the surface of the system. Following Deb et al.~\cite{Deb2017} we have predicted the exact values of the radii for different strange star candidates based on the observed values of the masses of the stellar objects and the chosen parametric values of the $\Lambda$ as well as the bag constant $\mathcal{B}$. The set of solutions satisfies all the physical requirements to represent strange stars. Interestingly, our study reveals that as the values of the $\Lambda$ and $\mathcal{B}$ increase the anisotropic system become gradually smaller in size turning the whole system into a more compact ultra-dense stellar object.
\end{abstract}

\keywords{General Relativity; metric potentials; strange stars}

\section{Introduction}\label{sec1}
Einstein's general theory of relativity (GR)~\cite{OR1996} represents a grand tool of gravitation to understand uniquely the fabric of the space-time and therefore the behavioral pattern of the cosmic bodies as well as phenomena. The first successful field theoretical application of GR to black hole solution by Schwarzschild~\cite{Schwarzschild1916} and to the structure of the universe by Einstein himself~\cite{Einstein1917} made pavement for a new avenue in the research fields of astrophysics and cosmology.

In the aspect of Cosmology, the cosmological constant $\Lambda$ as introduced by Einstein in general relativity to match the Mach principle and to have a non-expanding static solution of the universe, becomes very significant. But thirteen years later the pioneering observations by Edwin Hubble~\cite{Hubble1929} divulge that actually universe is expanding, which forced Einstein to abandon the concept of the cosmological constant as a `blunder'. Later, in 1998 scientists~\cite{Perlmutter1988,Riess1988} came up with some groundbreaking scientific evidence by observing the high redshift Type Ia supernova led to the decision that expansion of the universe is actually accelerating in manner, which reinstated the idea of Einstein's cosmological constant. Although the erstwhile cosmological constant was conjectured as a constant quantity but gradually it appears that $\Lambda$ is actually a scalar variable. This changes with time and the obtained decreasing value of $\Lambda$ ($\leq {10}^{-56}~{cm}^{-2}$)~\cite{Perlmutter1988,Riess1988} as obtained from the observational evidences agree with the concept of its variable nature. Now, as $\Lambda$ has been used in the field equations as a scalar variable, hence it should be a function of the spatial coordinate too. Importantly, likewise the framework of cosmology where variation of $\Lambda$ with the time has shown a significant effect, many researchers~\cite{Chen1990,Narlikar1991,Ray1993,Tiwari1996,Ray2008,HOSSEIN2012} have successfully studied the effect of the variation of $\Lambda$ in the realm of astrophysics with respect to the spatial coordinate. 

As a consequence of gravitational collapse we get a panorama of stellar formations which include white dwarfs, neutron stars and black holes (in the observational level) and also quark stars (in the theoretical level). In the present work, we are interested to study the strange stars, which are made of strange quark matter (SQM). The possibility of the existence of the hypothetical ultra-dense strange stars was first conjectured by several researchers~\cite{Itoh1970,Farhi1984,Alcock1986,Haensel1986}. The SQM, which is made of the equal number of up $(u)$, down $(d)$ and strange $(s)$ quarks, is assumed to be the true ground state for the confined hadrons~\cite{Farhi1984,Alcock1986,Alcock1988,Madsen1999}, as predicted by the strange quark matter hypothesis~\cite{Bodmer1971,Witten1984,Terazawa}. Though it is quite difficult to distinguish neutron stars from the strange stars in the basis of mass and radius but the strange stars, which are ultra-dense and massive stellar objects, have formed a different hypothetical branch for the compact stars. However, likewise the neutron stars and white dwarfs~\cite{Glendenning1995a,Glendenning1995b,Kettner1995}, strange stars do not exhibit continuum in the equilibrium configurations. Interestingly, the neutron star equation of state (EOS) failed to explain the compactness of the recently observed compact stellar objects like $4U~1820-30$, $SAX~J~1808.4-3658$, $4U~1728-34$, $Her~X-1$, $RX~J185635-3754$ and $PSR~0943+10$, etc., whereas SQM EOS~\cite{Alcock1986,Haensel1986,Weber2005,Perez-Garcia2010,Rodrigues2011,Bordbar2011} has satisfactorily explained the compactness of the stellar candidates.

The compact stars are basically considered as a spherically symmetric and isotropic ultradense stellar objects. However, isotropy maybe a favored feature but need not be a general characteristic of the stellar objects. Anisotropic factor ($\Delta=p_t-p_r$) which deals with the inhomogeneity in pressures into two components - radial pressure and tangential pressure, is actually a function to consider the internal situation of the star compared to the idealized isotropic case. The concept of anisotropy was originally proposed by Ruderman~\cite{Ruderman1972} and later on by several other scientists~\cite{Canuto1973,BL1974,HS1997}. As the reasons behind the pressure anisotropy different factors are thought to be responsible, such as very high density region in the core region, various condensate states (like pion condensates, meson condensates etc.), superfluid 3A, mixture of fluids of different types, rotational motion, presence of magnetic field, phase transition etc. The basic ideas involved in the anisotropy and their applications in the diverse fields are available in the recent literature~\cite{Ivanov2002,SM2003,MH2003,Usov2004,Varela2010,Rahaman2010,Rahaman2011,Rahaman2012,Kalam2012,Deb2015,Shee2016,Maurya2016,Maurya2017a,Deb2017,Ovalle2017a,Ovalle2017b,Isayev2017,Ivanov2017,Maurya2018,Estrada2018}.  

It is argued by Maurya~\cite{Maurya2017b} that the chances of having anisotropy is much higher in the compact stars because the interaction among the particles is highly relativistic, and they become very random to maintain any uniform distribution throughout the region. This relativistic nature of particles in the compact stars could be one of the possible reasons for giving birth of significant anisotropy in the compact stars. It is therefore understood that the anisotropic force inside the stellar systems makes the compact objects more compact than the isotropic condition which eventually makes the possible transition of a neutron star to strange star.

Under the above background, our motivation in the present paper is to study a spherically symmetric anisotropic stellar model and the effects of the cosmological constant, which is assumed to vary with the spatial coordinate, on the stellar system. The outline of this study is as follows: We provide the basic stellar equations to describe the assumed spherically symmetric anisotropic system and the Einstein field equations for the anisotropic stellar source in Sect.~\ref{sec2}. In Sect.~\ref{sec3} we have explained the necessity of the choice of appropriate EOS for SQM, whereas the solutions of the Einstein field equations and the expression for the different physical parameters has been featured in Sect.~\ref{sec4}. In Sect.~\ref{sec5} we have discussed junction conditions to derive the unknown constants. Several tests, viz., energy conditions, stability (via conservation equation, Herrera condition, and adiabatic index), compactification factor and redshift, have been conducted to study the physical validity of the stellar model in Sect.~\ref{sec6}. The last Sect.~\ref{sec7} is devoted as a platform for providing some salient features and concluding remarks along with a brief discussion.

\section{Basic stellar equations}\label{sec2}
To describe the spacetime of a compact stellar objects, we consider the spherically symmetric line element as 
\begin{equation}
ds^{2}=e^{\nu(r)}dt^{2}-e^{\lambda(r)}dr^{2}-r^{2}\left(d\theta^{2}+\sin^{2}\theta d\phi^{2} \right)\label{metric}.
\end{equation}

Literature survey~\cite{DL1998} shows that some interesting static spherically-symmetric perfect fluid solutions in canonical coordinates for the above metric were proposed by Tolman~\cite{Tolman1939}, Patwardhan and Vaidya~\cite{PV1943} and Mehra~\cite{Mehra1966} where $\lambda(r)=\ln(1+ar^2+br^4)$, and by Kuchowicz~\cite{Kuchowicz1968} and Leibovitz~\cite{Leibovitz1969} where $\nu(r)=Br^2+2\ln C$ with $a$, $b$, $B$ and $C$ as positive constants. The metric, therefore, with the mentioned metric potentials will be referred later on simply as the Tolman-Kuchowicz (TK) metric.

The Einstein field equations for the present anisotropic stellar system is given as
\begin{eqnarray}\label{Eeq}
G^i_j=R^i_j-\frac{1}{2}{g^i_j}R=-8\pi\left[T^{i(m)}_j+T^{i(vac)}_j\right],
\end{eqnarray}
where $T^{i(m)}_j$ and $T^{i(vac)}_j$ are the energy momentum tensor component for the anisotropic matter distribution and vacuum, respectively. The energy-momentum tensor components $T^{i(m)}_j$ and $T^{i(vac)}_j$ can be expressed in the following standard form:
\begin{eqnarray}
&\qquad T^{i(m)}_j=(\rho+p_t)u^i u_j-p_t g^i_j+(p_t-p_r)v^i v_j,\\
&\qquad T^{i(vac)}_j=\frac{1}{8\pi} g^i_j \Lambda(r),
\end{eqnarray}
where $p_r$, $p_t$ and $\rho$ represent the radial pressure, tangential pressure and matter-energy density, respectively. Here, $u^i$ is the 4-velocity and $v^i$ is the radial 4-vector of the fluid element.

We define the mass function of the present spherically symmetric stellar system as follows
\begin{equation}
m \left( r \right) =4\,\pi\,\int_{0}^{r}\!\rho \left( r \right) {r}^
{2}{dr}.\label{massfnc}
\end{equation}

We have considered the erstwhile Einsteinian cosmological constant as radially dependent, i.e. $\Lambda_r = \Lambda(r)$ and hence  the Einstein field equations for the metric (\ref{metric}) with energy-momentum tensor $T_{ij}$ are obtained as 
\begin{eqnarray} \label{rho}
 &\qquad \frac{1}{r^2}-e^{-\lambda}\left[\frac{1}{r^2}-\frac{\lambda'}{r}\right]=8\pi\rho+\Lambda_r, \\ \label{radialp}
&\qquad -\frac{1}{r^2}+e^{-\lambda}\left[\frac{1}{r^2}+\frac{\nu'}{r}\right]= 8\pi p_r-\Lambda_r, \\ \label{tangentialp}
&\qquad \frac{e^{-\lambda}}{2}\left[\nu''+\frac{\nu'^2-\lambda' \nu'}{2}+\frac{(\nu'-\lambda')}{r}\right]= 8\pi p_t-\Lambda_r,  
\end{eqnarray}
where $`\prime$' denotes the differentiation with respect to the radial coordinate $r$. In above field equations we have considered geometrized units $G = c = 1$.

Now, the equation of continuity, $T^i_{j;i}=0$ of the present anisotropic system reads
\begin{eqnarray}\label{TOVeq}
\frac{1}{2}\,\nu'\,(\rho+p_r)+\frac{d}{dr}\left(p_r-\frac{\Lambda_r}{8\pi}\right)+\frac{2}{r}(p_r-p_t)=0.
\end{eqnarray}

\section{Equation of state}\label{sec3}
Now to describe the interior structure of the stellar system we assume the radial pressure is directly related to the energy density by a relation, known as the equation of state (EOS). Once we define EOS, immediately the Einstein field equations~(6)-(8) can easily be tackled. In the present investigation, we have assumed the MIT bag EOS to describe anisotropic SQM, made of up, down and strange quark only~\cite{Witten1984}. Many researchers have successfully used the MIT bag EOS in their recent studies~\cite{1,2,3,4,5,6,7,8} for SQM. The MIT bag EOS is given as follows
\begin{equation*}
p_r=\alpha \left(\rho - 4\mathcal{B}\right), \label{EOS1}
\end{equation*}
where $\alpha$ is a constant and it has value $1/3$ for the massless strange quarks whereas it is equal to $0.28$ for the massive strange quarks having mass $250~MeV$~\cite{Stergioulas2003}. Here $\mathcal{B}$ represents the bag constant. It is worth mentioning that to include all the corrections due to the energy and pressure functions of SQM, in the simplified and linear MIT bag model EOS an adhoc bag function has been introduced. In the present study, the mathematical analysis has been carried out by considering $\alpha=0.28$ and $B=60~MeV/{fm}^3$.

\section{Solution of the Einstein field equations}\label{sec4}
\begin{figure}[!htp]
\centering
\includegraphics[width=6cm]{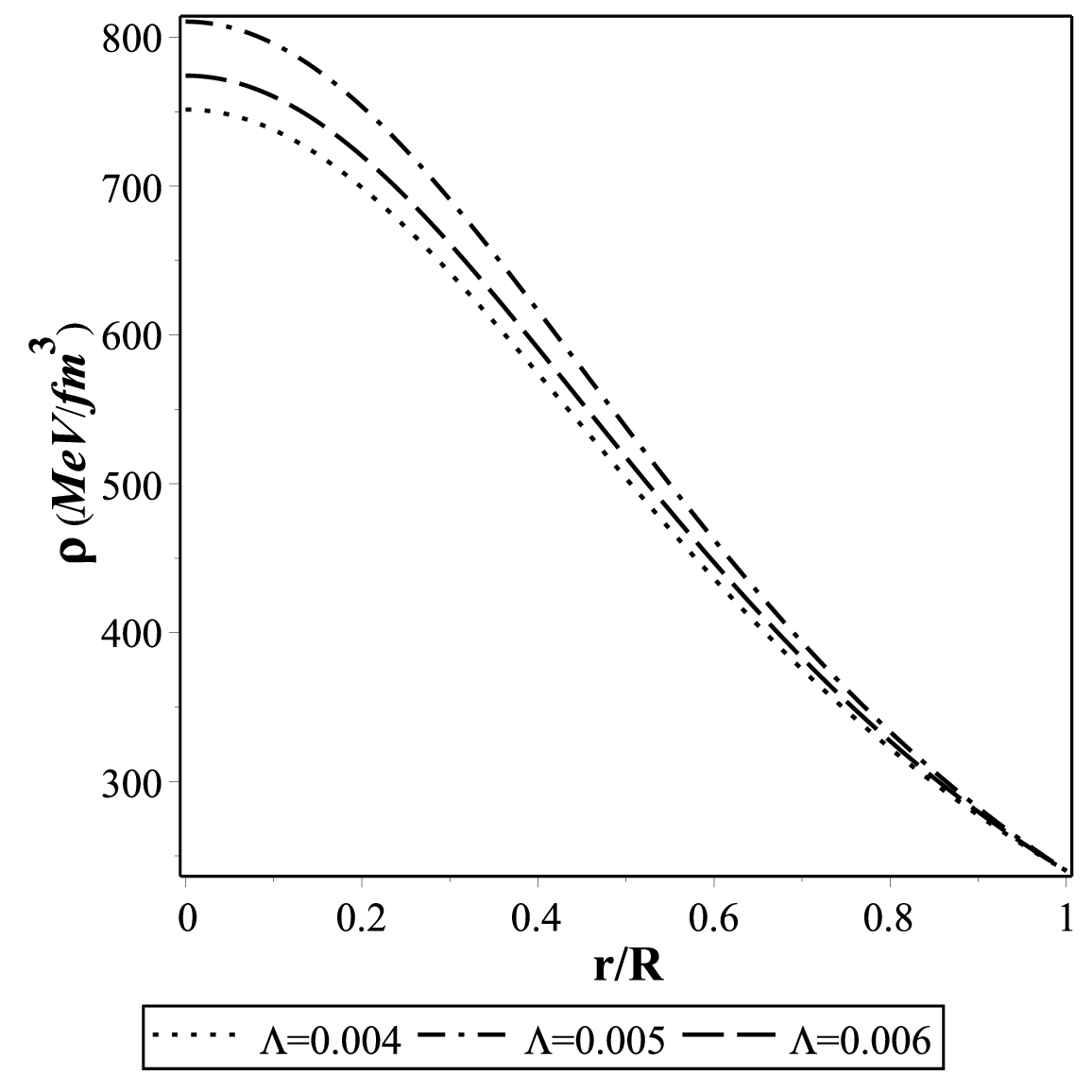}
\caption{Behavior of density as a function of the fractional radial distance for the strange star candidate $LMC~X-4$ is shown. Here and in what follows $B=60~MeV/{fm}^3$}\label{Fig1}
\end{figure}

\begin{figure}[!htp]
\centering
\subfloat{\includegraphics[width=5cm]{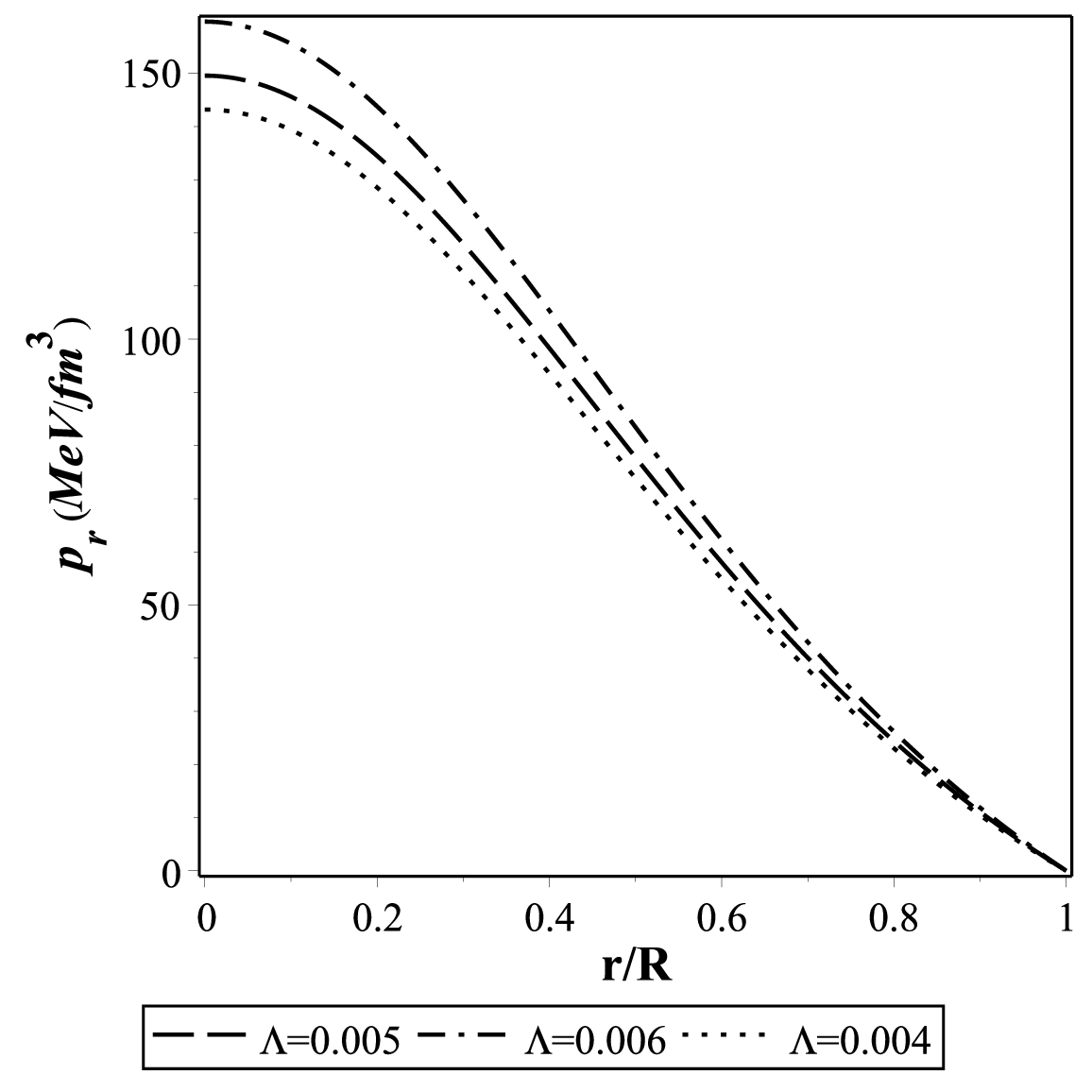}}
\
\subfloat{\includegraphics[width=5cm]{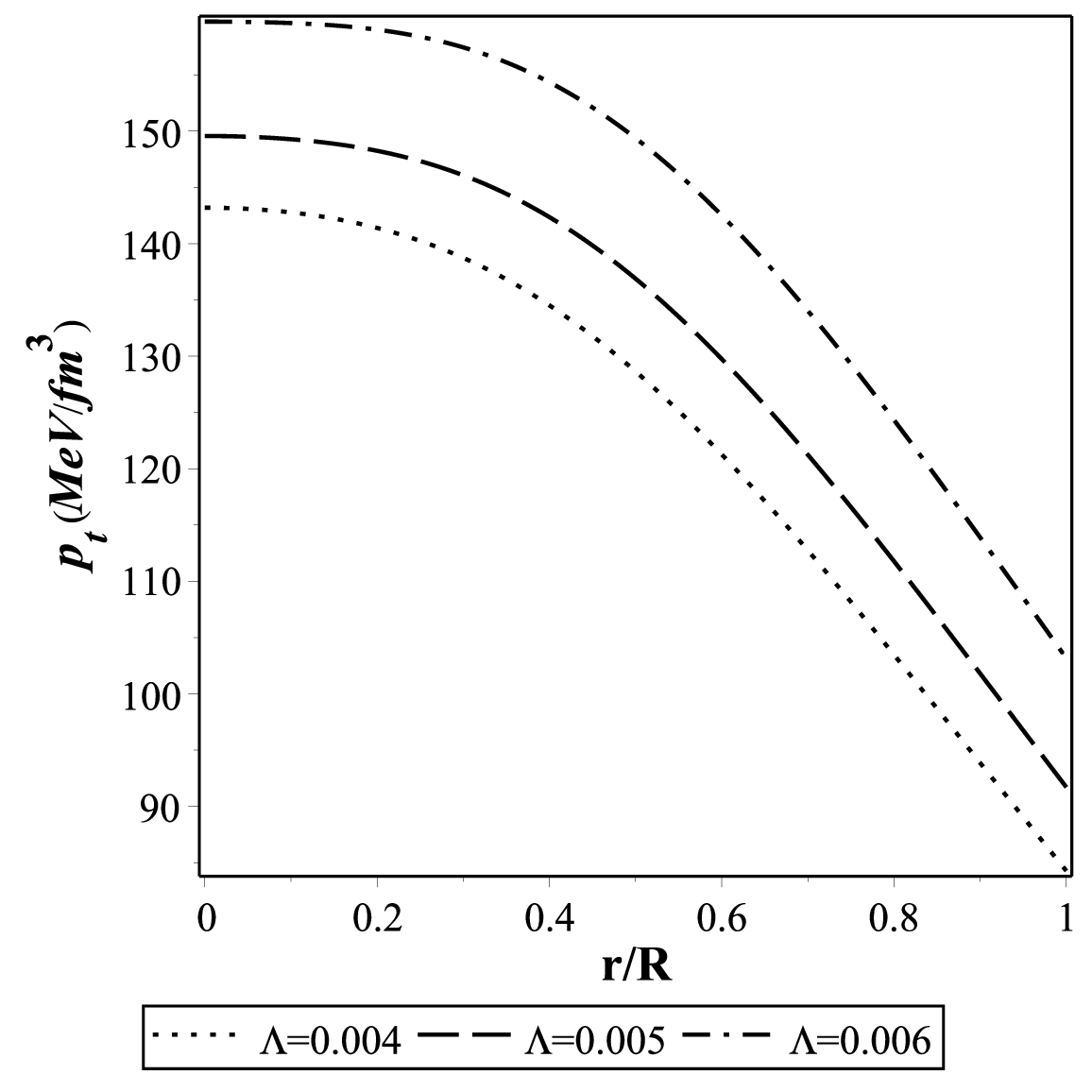}}
\caption{Behavior of the radial pressure $p_r$ (left panel) and the tangential pressure $p_t$ (right panel) as a function of the fractional radial distance for the strange star candidate $LMC~X-4$ are shown}\label{Fig2}
\end{figure}

\begin{figure}[!htp]
\centering
\includegraphics[width=5cm]{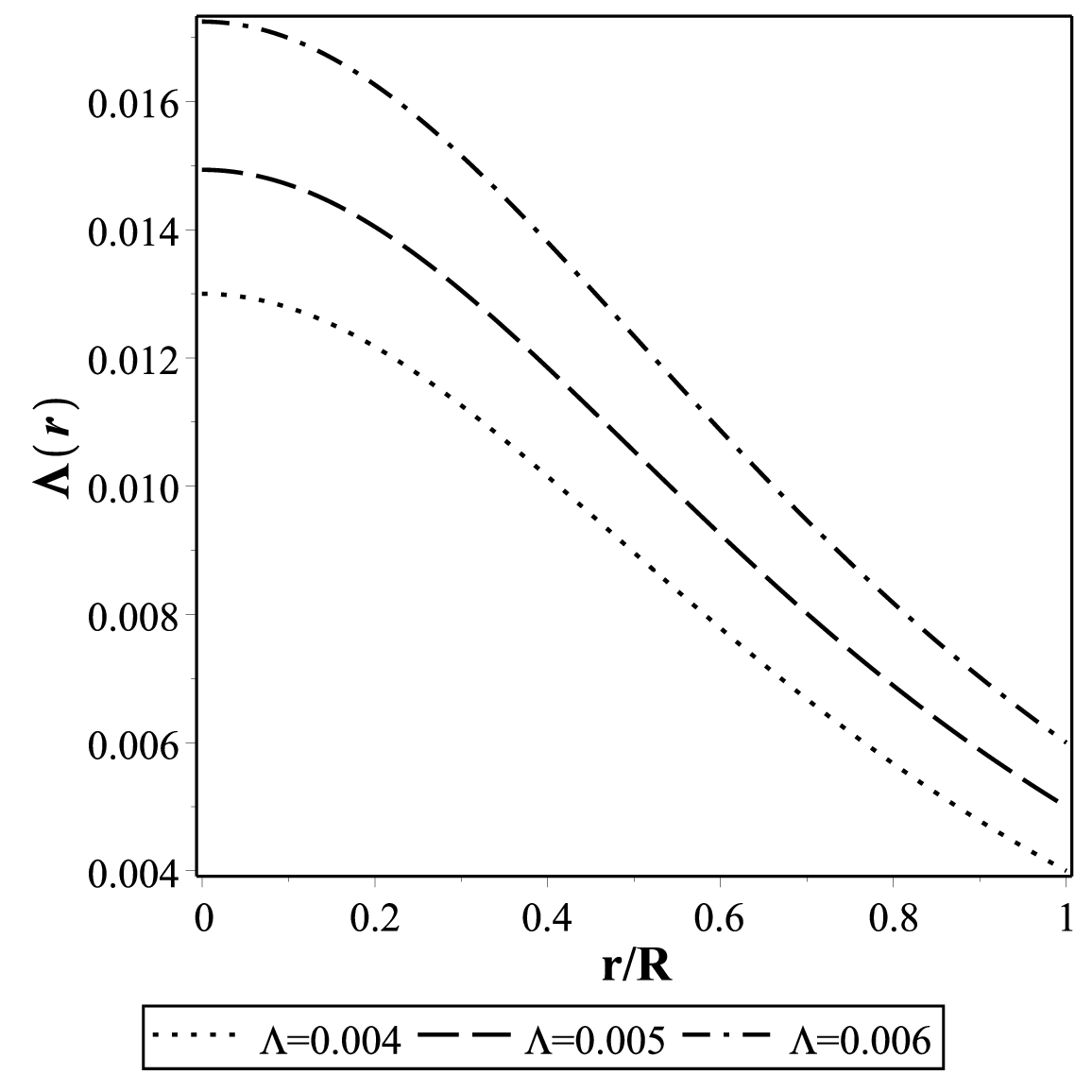}
\caption{Behavior of the cosmological constant $\Lambda(r)$ as a function of the fractional radial distance for the compact star $LMC~X-4$ is shown}\label{Figcosmo}
\end{figure}

By using the metric (\ref{metric}) and Eqs. (\ref{rho})-(\ref{tangentialp}) and (\ref{EOS1}), we can write the expressions for radial pressure $p_r$, tangential pressure $p_t$ and energy density $\rho$ in the following form as
\begin{eqnarray}\label{rho1}
&\qquad\hspace{-2cm} \rho=\frac{2 [a +2 b r^2 + B(1 + ar^2 + br^4)]+\beta(1 + ar^2 + br^4)^2}{ 8\pi(1+\alpha)(1 + ar^2 + br^4)^2},\\ \label{radialp1}
&\qquad\hspace{-2cm}  p_r=\frac{2\alpha [a +2 b r^2 + B(1 + ar^2 + br^4)]-\beta(1 + ar^2 + br^4)^2}{ 8\pi(1+\alpha)(1 + ar^2 + br^4)^2},\\ \label{tangentialp1}
&\qquad\hspace{-0.8cm}  p_t=\Big\lbrace\frac{[2 B + a^2 r^2 + b r^2 (3 + b r^4) +a (2 + B r^2 + 2 b r^4 + B^2 r^4) + B^2r^2 (1 + b r^4)]} {8\pi(1 + ar^2 + br^4)^2} \nonumber \\
&\qquad\hspace{-0.5cm} - \frac{2 [a + 2 b r^2+ B(1 + a r^2 + b r^4)] + \beta (1 + a r^2 + b r^4)^2}{8\pi(1+\alpha)(1 + ar^2 + br^4)^2}\Big\rbrace,\\ \label{Lambda}
&\qquad\hspace{-0.6cm} \Lambda_r=\Big[\frac{(1+\alpha)[a^2 r^2 +b r^2 (5 + b r^4) + a (3 + 2 b r^4)]-2 B(1+ a r^2 + b r^4)}{(1 + \alpha)(1 + a r^2 + b r^4)^2} \nonumber \\
&\qquad\hspace{0cm} -\frac{2(a+ 2 b r^2)+\beta(1 + a r^2 + b r^4)^2} {(1+\alpha)(1 + a r^2 + b r^4)^2}\Big]. 
\end{eqnarray}
where $\beta=32\,\mathcal{B}\pi \,\alpha$. We have shown behaviour of the density function in Fig.~\ref{Fig1}. Again, the variation of the pressures viz., $p_r$ and $p_t$ against the radial coordinate have been featured in Fig.~\ref{Fig2}. From Figs.~\ref{Fig1} and \ref{Fig2} we find $\rho$, $p_r$ and $p_t$ decrease monotonically from the maximum value at the centre to the minimum value at the surface and confirm regularity of the achieved solutions. We have presented variation of the varying cosmological constant against the radial coordinate in Fig.~\ref{Figcosmo}

Hence, the anisotropic stress $\Delta(r)=p_t-p_r$ for the present system is as follows
\begin{eqnarray}
 &\qquad\hspace{-0.5cm} \Delta=\frac{r^2 [a^2 + B^2 + b^2 r^4 + a (-B + 2 b r^2 + B^2 r^2) - b (1 + 2 B r^2 - B^2 r^4)]} {8\pi(1 + a r^2 + b r^4)^2}. \nonumber \\
\end{eqnarray}

\begin{figure}[!htp]
\centering
\includegraphics[width=5cm]{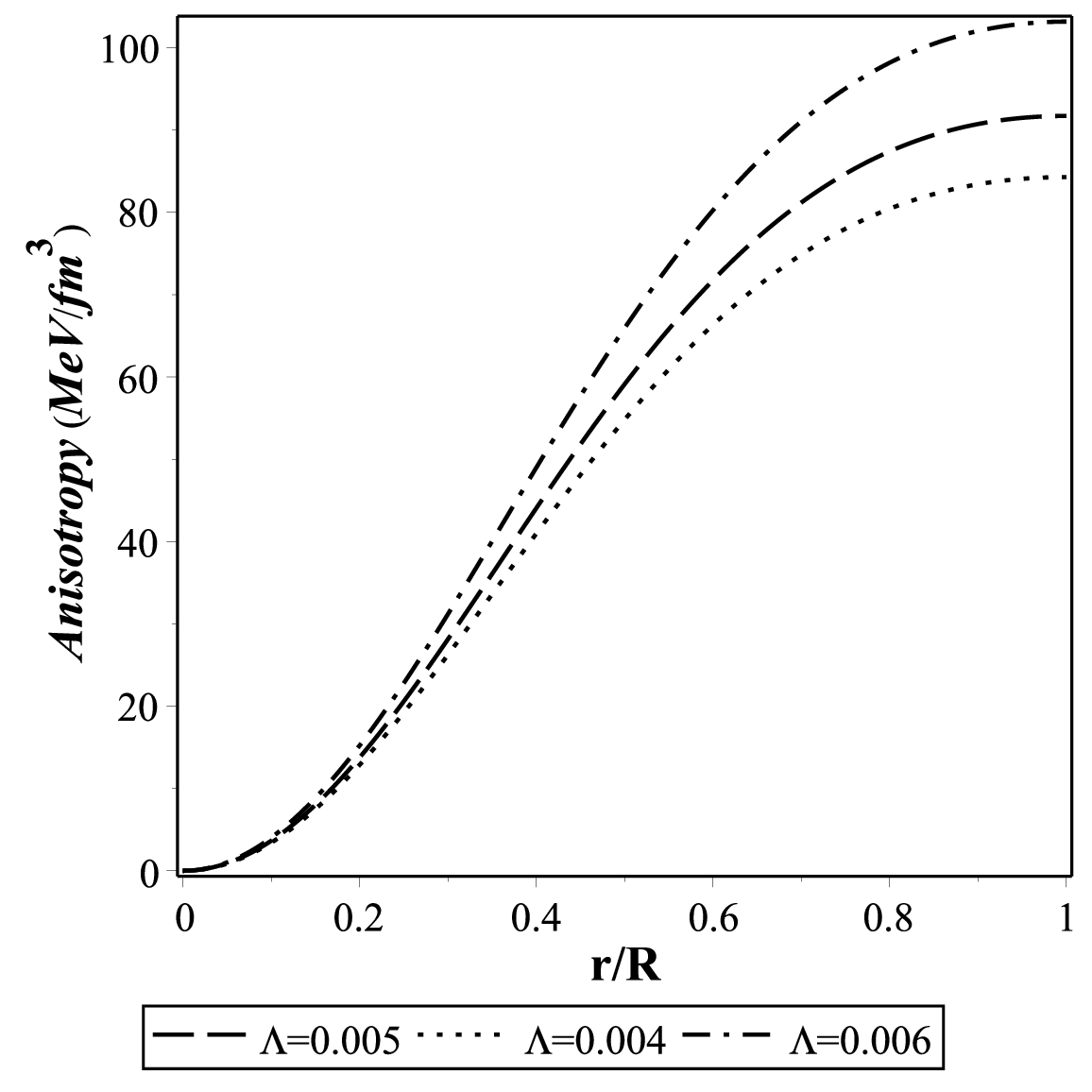}
\caption{Behavior of the anisotropic stress $\Delta$ as a function of the fractional radial distance for the strange star candidate $LMC~X-4$ is shown}\label{Fig3}
\end{figure}

We have shown the variation of the anisotropic stress $\Delta$ with respect to the fractional radial coordinate $r/R$ in Fig.~\ref{Fig3}.

\section{Junction conditions}\label{sec5}
The first fundamental form of the boundary surface involved in the metric~(\ref{metric}) should be the same whether obtained from the interior or exterior space-time and this guarantees that for some coordinate system the metric components $g_{ij}$ will be continuous across the surface. Therefore, we consider that the exterior system is equivalent to the metric 
\begin{eqnarray}
&\qquad\hspace{-1cm} ds^2= \left(1-\frac{2M}{r}-\frac{\Lambda r^2}{3}\right)dt^2- \left(1-\frac{2M}{r}-\frac{\Lambda r^2}{3}\right)^{-1} dr^2  -r^2 (d \theta^2 +\sin ^2 \theta d \phi ^2),
\end{eqnarray}
which is the modified Schwarzschild metric with the cosmological constant and hence turns into the usual Schwarzschild metric for vanishing $\Lambda$.
   
The continuity of the metric functions, involved in the interior and the exterior metrics, at the outer boundary of the fluid sphere ($r=R$), as well as the requirement of matching condition for the radial pressure, i.e. $p_r = 0$ at the surface, immediately provides the constants $a$, $b$, $B$ and $C$ in the following forms
\begin{eqnarray}
&\qquad\hspace{-1cm} B={\frac {2\Lambda {R}^{3}-6M}{2{R}^{2} \left( \Lambda_{{
1}}{R}^{3}+6M-3R \right) }}, \\
&\qquad\hspace{-1cm} C={{\rm e}^{{\frac {\left( \Lambda {R}^{3}+6M-3R \right) \ln  \left(1-{\frac {2M}{R}}-\frac{1}{3}\Lambda {R}^{2}\right) -\Lambda {R}^{3}+3M}{2\Lambda {R}^{3}+12M-6R}  }}},\\
&\qquad\hspace{-0.5cm} a={\frac {-4{R}^{6}\alpha{\Lambda}^{2}+ \left( 6
\alpha \Lambda -9\beta \right) {R}^{4}-48 M{R}^{3}\alpha 
\Lambda +90 MR\alpha-144 {M}^{2}\alpha}{2{R}^{2}\alpha \left( 
\Lambda {R}^{3}+6 M-3 R \right) ^{2}}},\nonumber \\ \\
&\qquad\hspace{-1cm} b={\frac { \left( 2{R}^{6}{\Lambda}^{2}+24 M{R}^{3}
\Lambda +72{M}^{2}-54MR \right) \alpha+9\beta {R}^{4}}{
2\alpha {R}^{4} \left( \Lambda {R}^{3}+6 M-3R \right) ^{2}}}.
\end{eqnarray}

\begin{figure}[!htp]
\centering
\includegraphics[width=5cm]{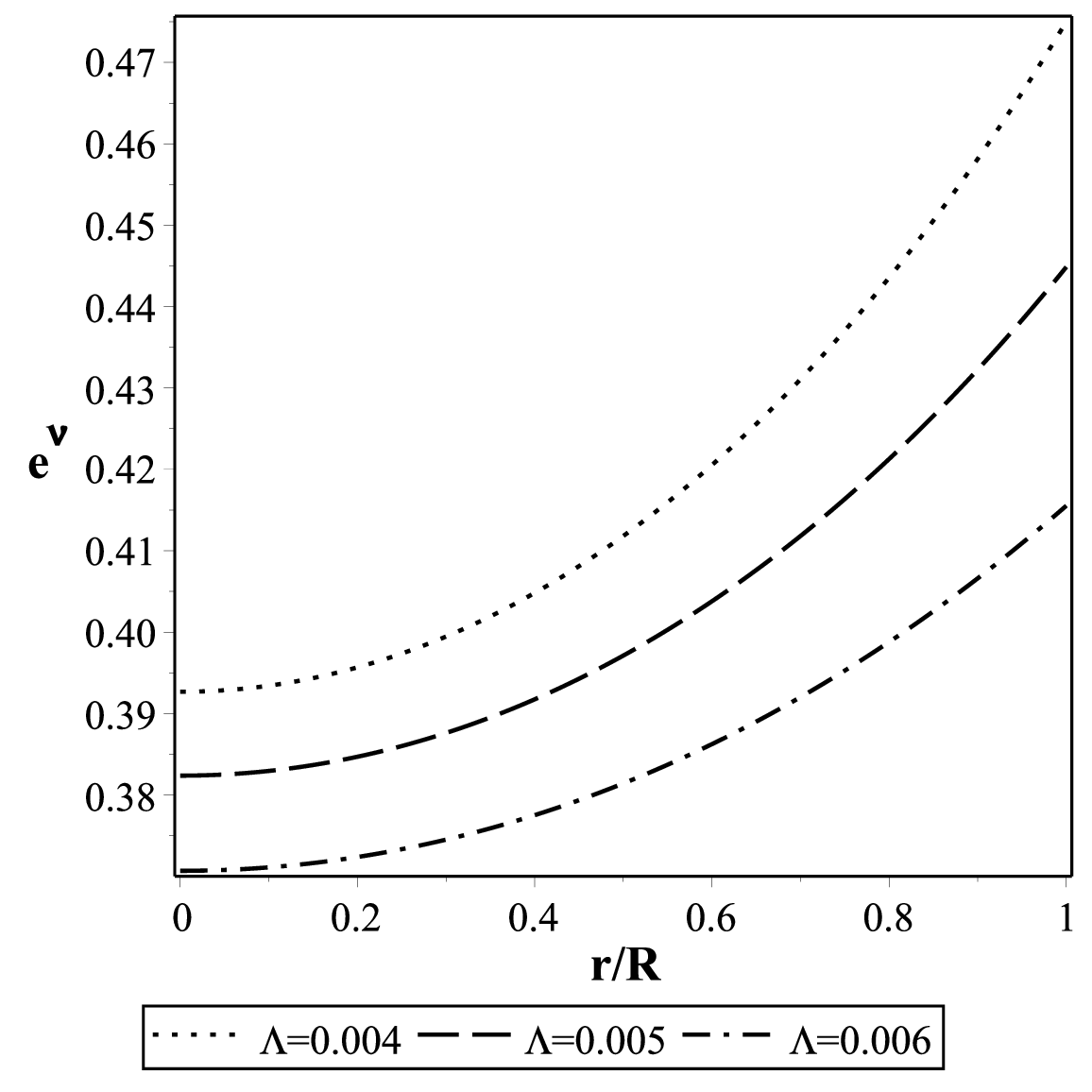}
\includegraphics[width=5cm]{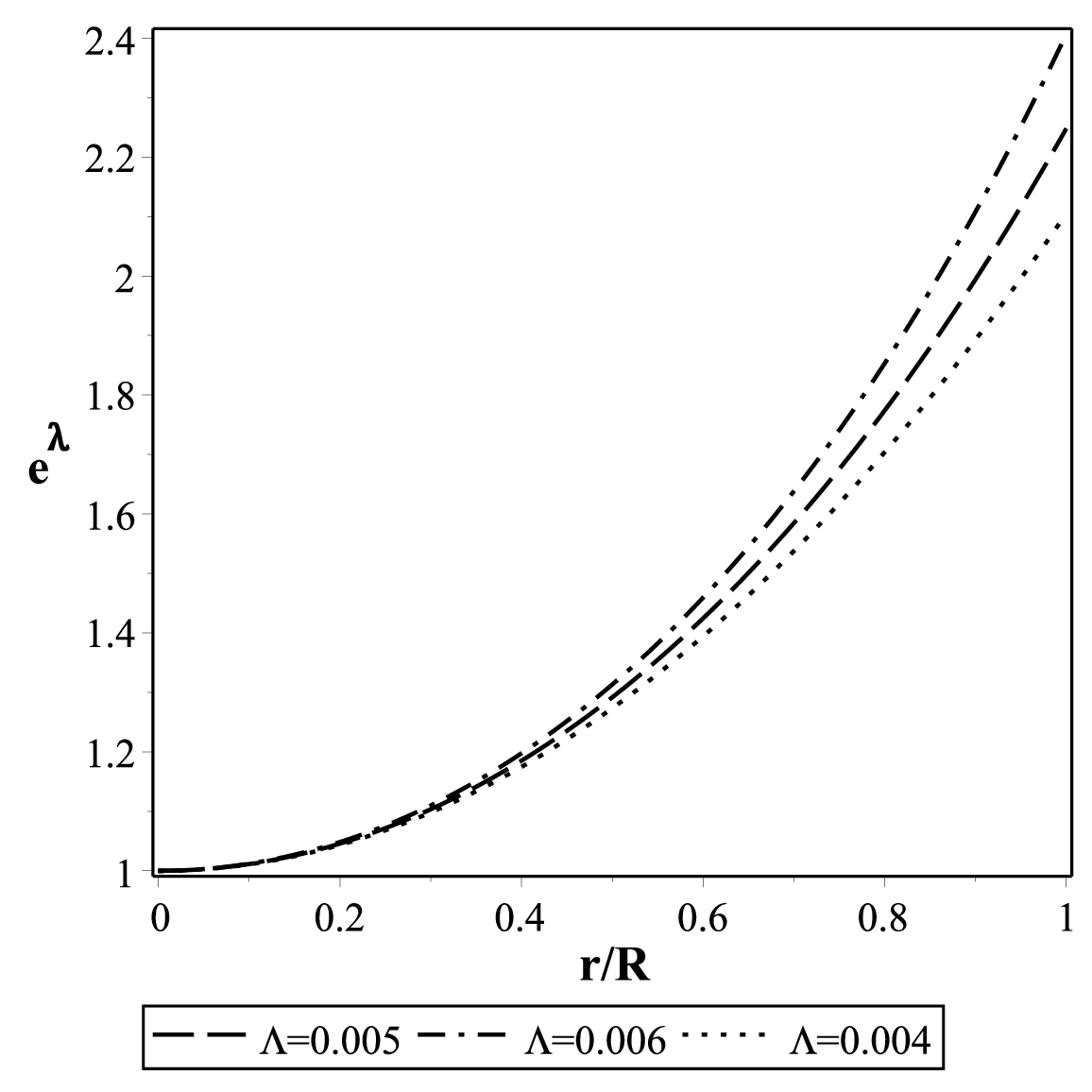}
\caption{Behavior of the metric potentials viz., $\rm e^\nu$ (left panel) and $\rm e^\lambda$ (right panel) as a function of the fractional radial distance for the strange star candidate $LMC~X-4$ are shown}\label{Figpot}
\end{figure}

In the later part of the article it can be observed that tuning of these constants are very important to determine different physical properties of compact stars. The variations of the metric potentials viz., $\rm e^\nu$ and $\rm e^{\lambda}$ based on the obtained values of the constants, against the radial coordinate is shown in Fig.~\ref{Figpot}.

Now, many scientists~\cite{Maharaj2012,Takisa2013} have shown that anisotropy is minimum at the surface, on the other hand, many others~\cite{Harko2002,Kalam2012,HOSSEIN2012} have also shown that anisotropy is maximum at the surface. To resolve the confusion that what should be the real feature of the anisotropic stress at the surface Deb et al.~\cite{Deb2017} have maximized anisotropy at the surface without pre-assuming it as maximum or minimum at $r=R$ and conclusively showed that for a stable and spherically symmetric anisotropic stellar system the anisotropic stress must be maximum at the surface as an inherent property of the compact stars. In the present study, following Deb et al.~\cite{Deb2017} we have predicted different physical parameters for the strange star candidates, including the exact values of radius by considering the observed values of mass of the strange star candidates for $\mathcal{B}=60~MeV/{fm}^3$ (see Table.~\ref{Table 1}).


\begin{table*}
  \centering
    \caption{Numerical values of physical parameters for the different star candidates for $\Lambda=0.005$ and $\mathcal{B}=60 MeV/{fm}^3$ } \label{Table 1}
        \scalebox{0.9}{
\begin{tabular}{ ccccccccccccccccccccccccccc}
\hhline{=========}
Strange & Observed  & Predicted & ${{\rho}_{c}}$  & ${{p}_{c}}$ & Surface & $\frac{2M}{R}$ \\ 
 Stars & Mass ($M_{\odot}$) & Radius (Km) & $(gm/{cm}^3)$ & $(dyne/{cm}^2)$ & Redshift &\\
\hline

$\hspace{-0.2cm}PSR~{J1614-2230}$ & $1.97 \pm 0.04$~\cite{demorest} & $9.949 \pm 0.016$ & $1.973 \times {{10}^{15}}$ & $3.888 \times {{10}^{35}}$ & $0.58$ & $0.54$ \\ 
 
$Vela~X-1$ & $1.77 \pm 0.08 $~\cite{dey2013} & $9.824 \pm 0.067$ & $1.725\times {{10}^{15}}$ & $3.265 \times {{10}^{35}}$ & $0.53$ & $0.46$ \\ 

$4U~1608-52$ & $1.74 \pm 0.14$~\cite{guver2010a} & $9.798 \pm 0.126$ & $1.695\times {{10}^{15}}$ & $3.188 \times {{10}^{35}}$ & $0.52$ & $0.44$ \\ 

$PSR~J1903~+~327$ & $1.667 \pm 0.021$~\cite{dey2013}  & $9.727 \pm 0.022$ & $1.625\times {{10}^{15}}$ & $3.013 \times {{10}^{35}}$ & $0.51$ & $0.43$ \\ 

$4U~1820-30$ & $1.58 \pm 0.06$~\cite{guver2010b} & $9.628 \pm 0.074$ & $1.552\times {{10}^{15}}$ & $2.828 \times {{10}^{35}}$ & $0.48$ & $0.39$ \\ 

$Cen~X-3$ & $1.49 \pm 0.08$~\cite{dey2013} & $9.508 \pm 0.115$ & $1.487\times {{10}^{15}}$ & $2.665 \times {{10}^{35}}$ & $0.46$ & $0.36$ \\ 

$EXO~{1785-248}$ & $1.3 \pm 0.2$~\cite{ozel2009} & $9.189 \pm 0.396$ & $1.385\times {{10}^{15}}$ & $2.408 \times {{10}^{35}}$ & $0.42$ & $0.31$ \\ 

$LMC~X - 4$ & $1.29 \pm  0.05$~\cite{dey2013} & $9.170 \pm 0.098$ & $1.380\times {{10}^{15}}$ & $2.396 \times {{10}^{35}}$ & $0.41$ & $0.30$ \\

\hhline{=========} 
\end{tabular}  }
  \end{table*}


\section{The physical properties of the stellar model}\label{sec6}
In this section we will discuss physical validity of the achieved solutions based on the few physical tests viz., energy conditions, equilibrium of forces, Herrera cracking concept, etc.

\subsection{Energy conditions}
The energy conditions, viz., Null Energy Condition (NEC), Weak Energy Condition (WEC) and Strong Energy
Condition (SEC) will be valid only when the following inequalities hold simultaneously for our system given as
\begin{eqnarray}\label{eq21}
&\qquad\hspace{-0.5cm}~NEC:\rho+p_r\geq 0,~\rho+p_t\geq 0, \\ \label{eq22}
&\qquad\hspace{-0.5cm}~WEC: \rho+p_r\geq 0,~\rho\geq 0,~\rho+p_t\geq 0, \\ \label{eq23}
&\qquad\hspace{-0.5cm}~SEC: \rho+p_r\geq 0,~\rho+p_r+2\,p_t\geq 0, \\ \label{eqdec}
&\qquad\hspace{-0.5cm}~DEC: {{\rho}}\geq 0,~{{\rho}}-{{p_r}}\geq 0,~{{\rho}}-{{p_t}}\geq 0.
\end{eqnarray}

\begin{figure}[!htp]
\centering
\subfloat{\includegraphics[width=4cm]{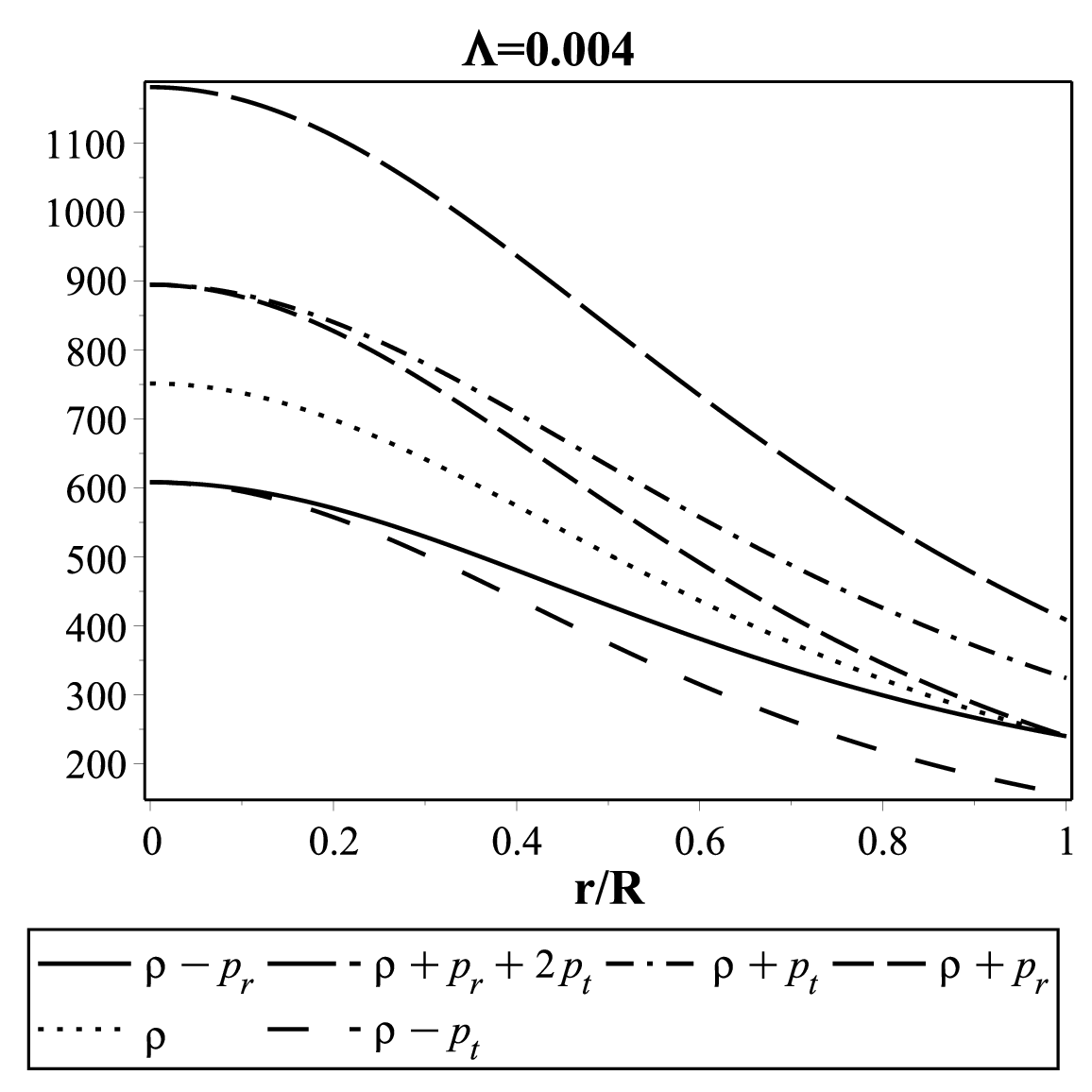}}
\
\subfloat{\includegraphics[width=4cm]{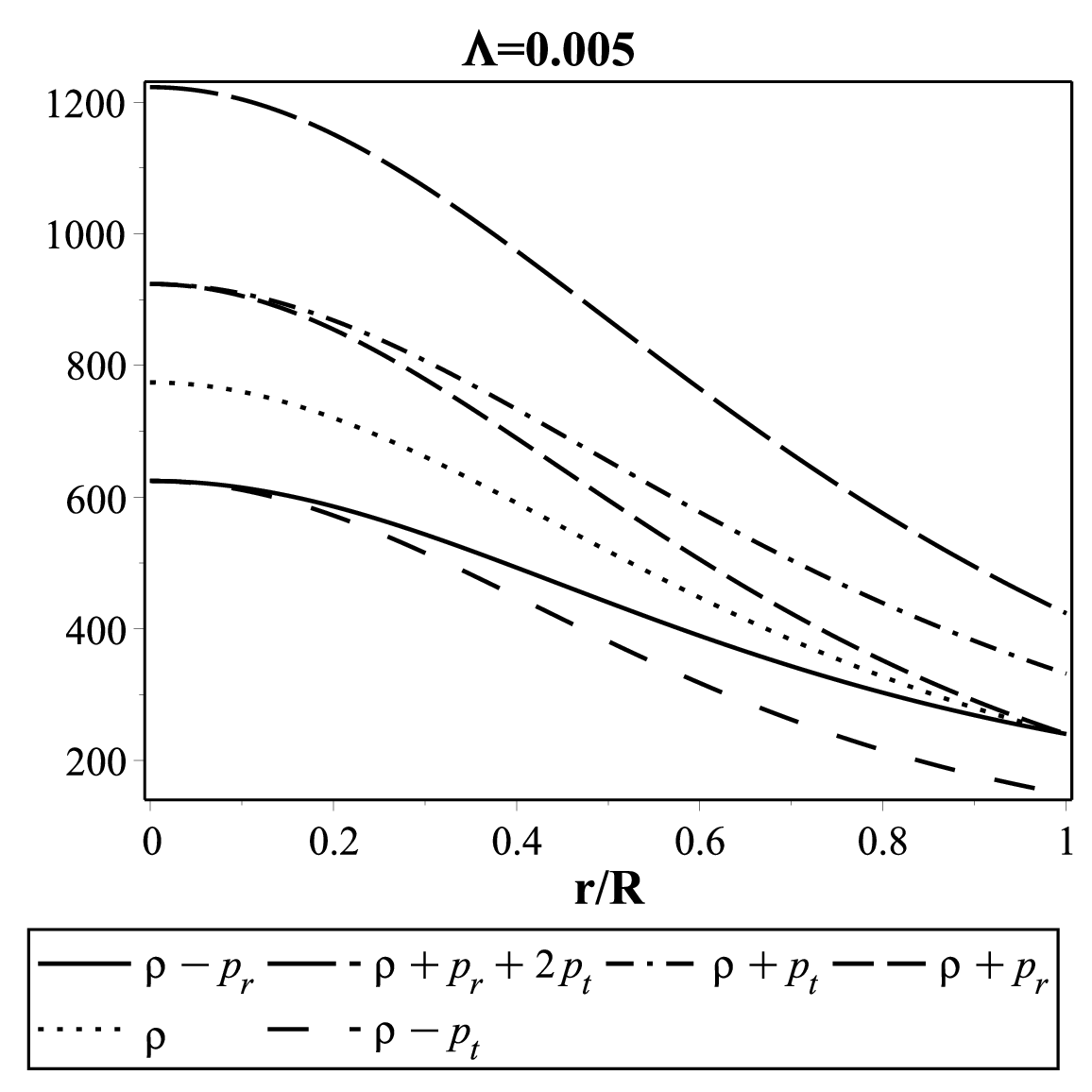}}
\
\subfloat{\includegraphics[width=4cm]{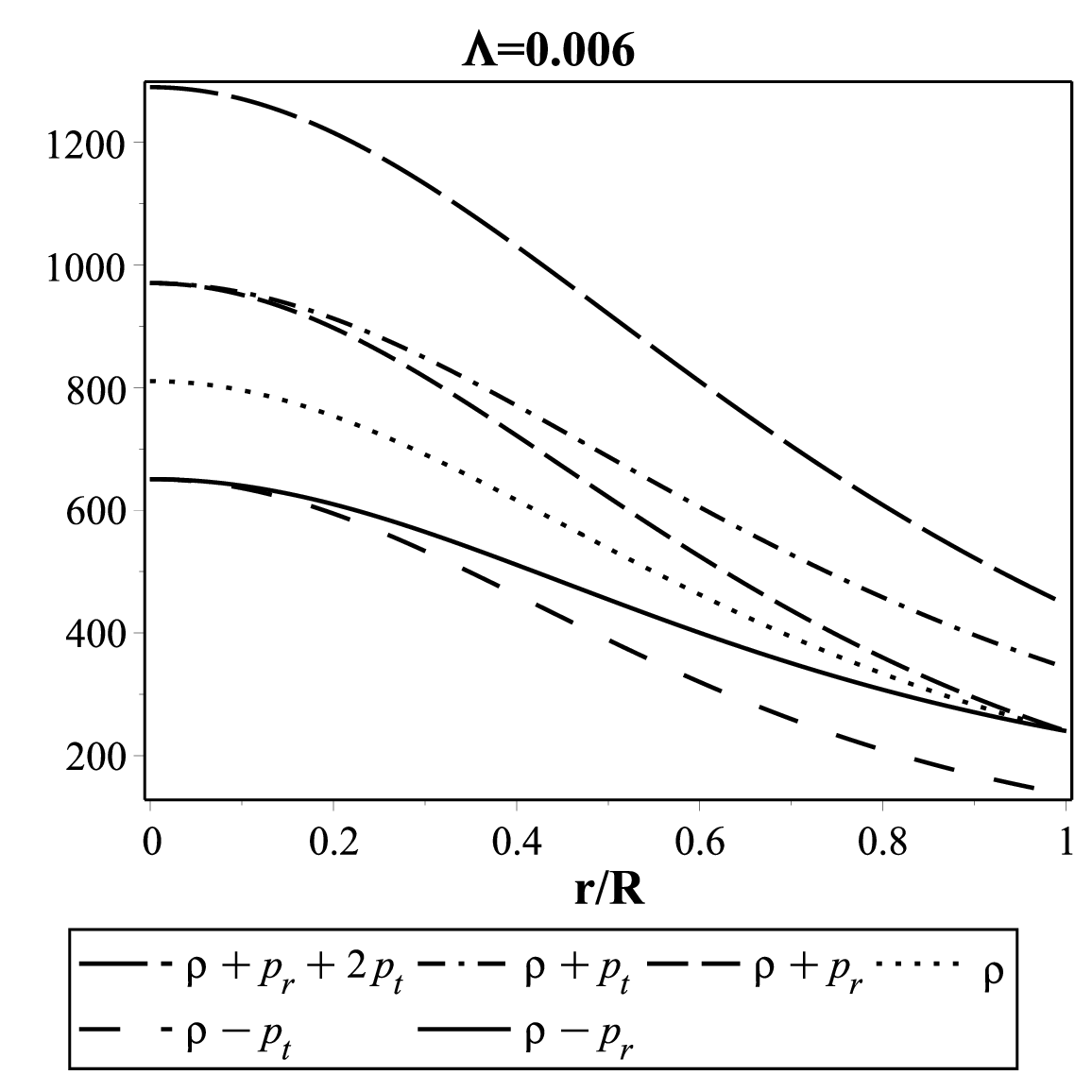}}
\caption{Behavior of the energy conditions for $\Lambda=0.004$ (left panel), $\Lambda=0.005$ (middle panel) and $\Lambda=0.006$ (right panel) as a function of the fractional radial distance for the strange star candidate $LMC~X-4$ are shown}\label{FigEC}
\end{figure}

The energy conditions are plotted in Fig.~\ref{FigEC} which features that as our system is consistent with all the inequalities simultaneously, it confirms that the achieved solution is physically viable.    

\subsection{Stability of the stellar model}

\subsubsection{TOV equation}
The Tolaman-Oppenhiemer-Volkoff (TOV)~\cite{Tolman1939,OV1939} equation for the anisotropic matter with variable cosmological
constant ($\Lambda_r$) in radial direction are given as
\begin{eqnarray}
\frac{1}{2}\,\nu'\,(\rho+p_r)+\frac{d}{dr}\left(p_r-\frac{\Lambda_r}{8\pi}\right)+\frac{2}{r}(p_r-p_t)=0.
\end{eqnarray}

It is Obviously that the above generalized TOV equation describes the equilibrium
condition for the strange star subject to the gravitational ($F_g$) and hydrostatic ($F_h$) plus
another anisotropic force ($F_a$) due to the anisotropic nature of the stellar object. Now, the above
equation can be written via different forces as
\begin{eqnarray}
F_g+F_h+F_a=0,
\end{eqnarray}
with
\begin{eqnarray}
&\qquad\hspace{-5cm}F_g=-\frac{1}{2}\,\nu'\,(\rho+p_r)\nonumber \\
& \qquad\hspace{-2.5cm} =- \frac{2Br[a + 2 b r^2 + B (1 + a r^2 + b r^4)]}{8\pi(1 + a r^2 + b r^4)^2},\\
&\qquad\hspace{-5cm} F_h=-\frac{d}{dr}\left(p_r-\frac{\Lambda_r}{8\pi}\right)\nonumber \\
& \qquad\hspace{-1.8cm} =\frac{2r[-a^2 + 2 a (B - b r^2) + b (1 + 4 B r^2 - b r^4)]}{8\pi(1 + a r^2 + b r^4)^2},\\
&\qquad\hspace{-5.7cm} F_a=\frac{2}{r}(p_t-p_r)\nonumber \\
& \qquad=\frac{2r [a^2 + B^2 + b^2 r^4 + a (-B + 2 b r^2 + B^2 r^2) - b (1 + 2 B r^2 - B^2 r^4)]} {8\pi(1 + a r^2 + b r^4)^2}.\nonumber \\
\end{eqnarray}

\begin{figure}[!htp]
\centering
\subfloat{\includegraphics[width=4cm]{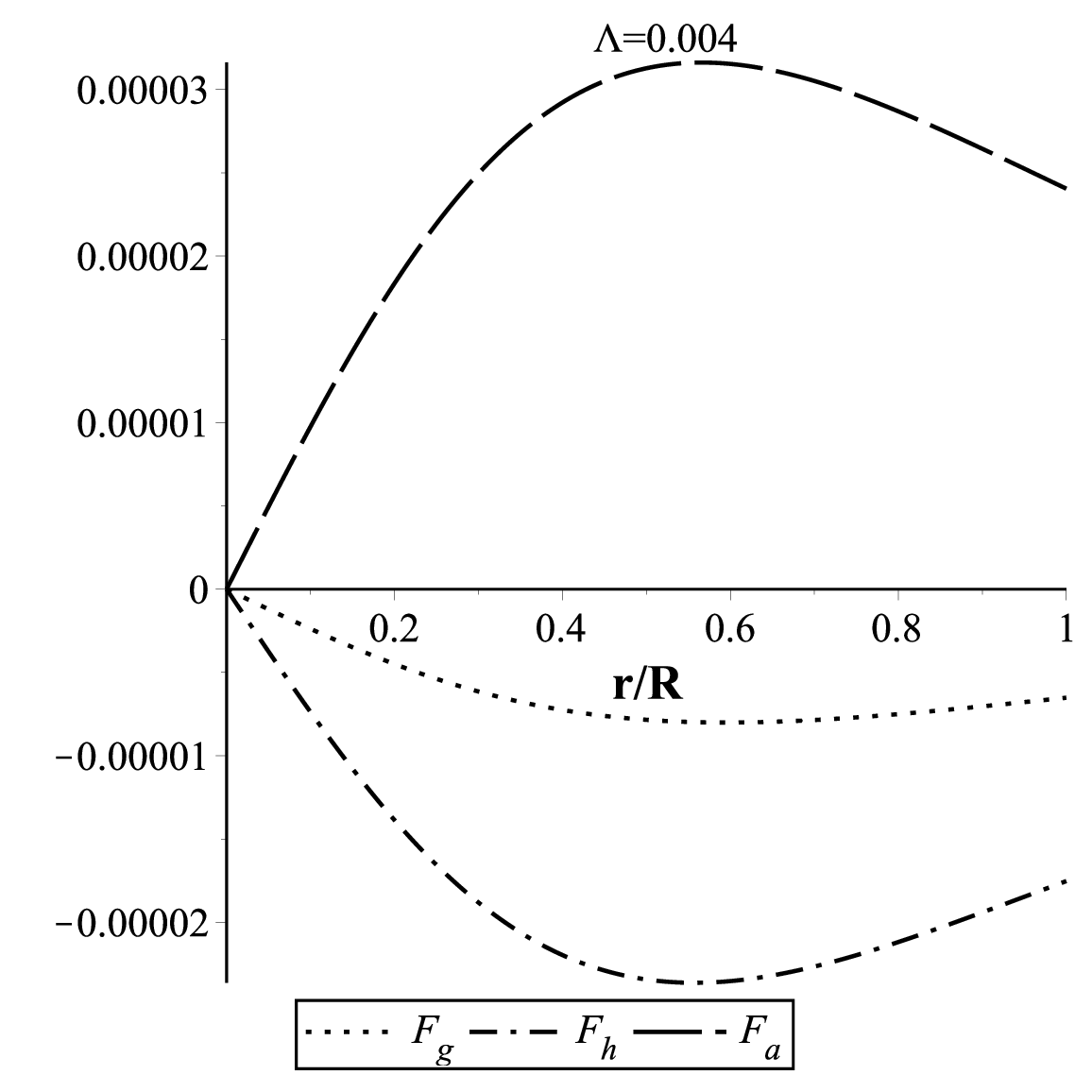}}
\
\subfloat{\includegraphics[width=4cm]{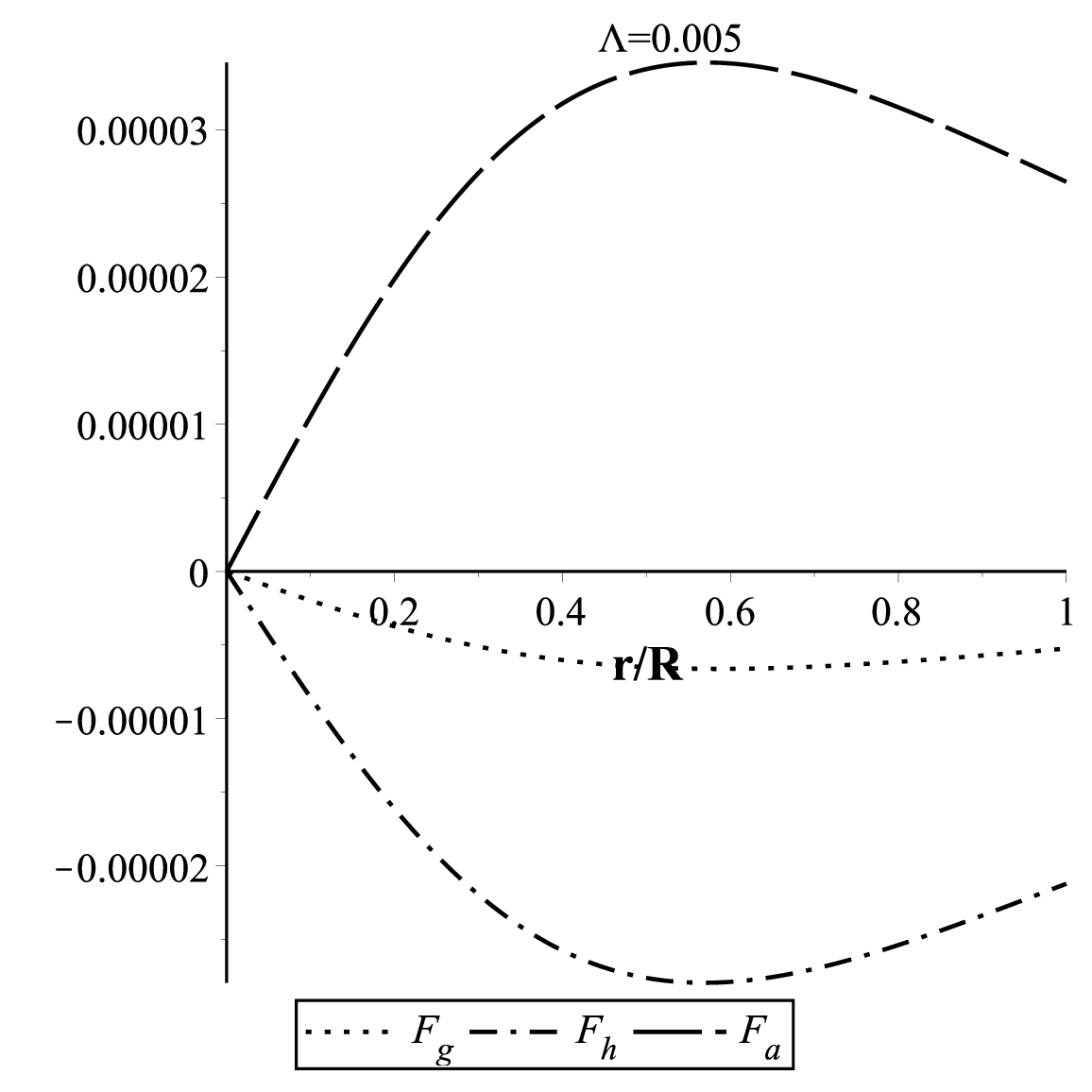}}
\
\subfloat{\includegraphics[width=4cm]{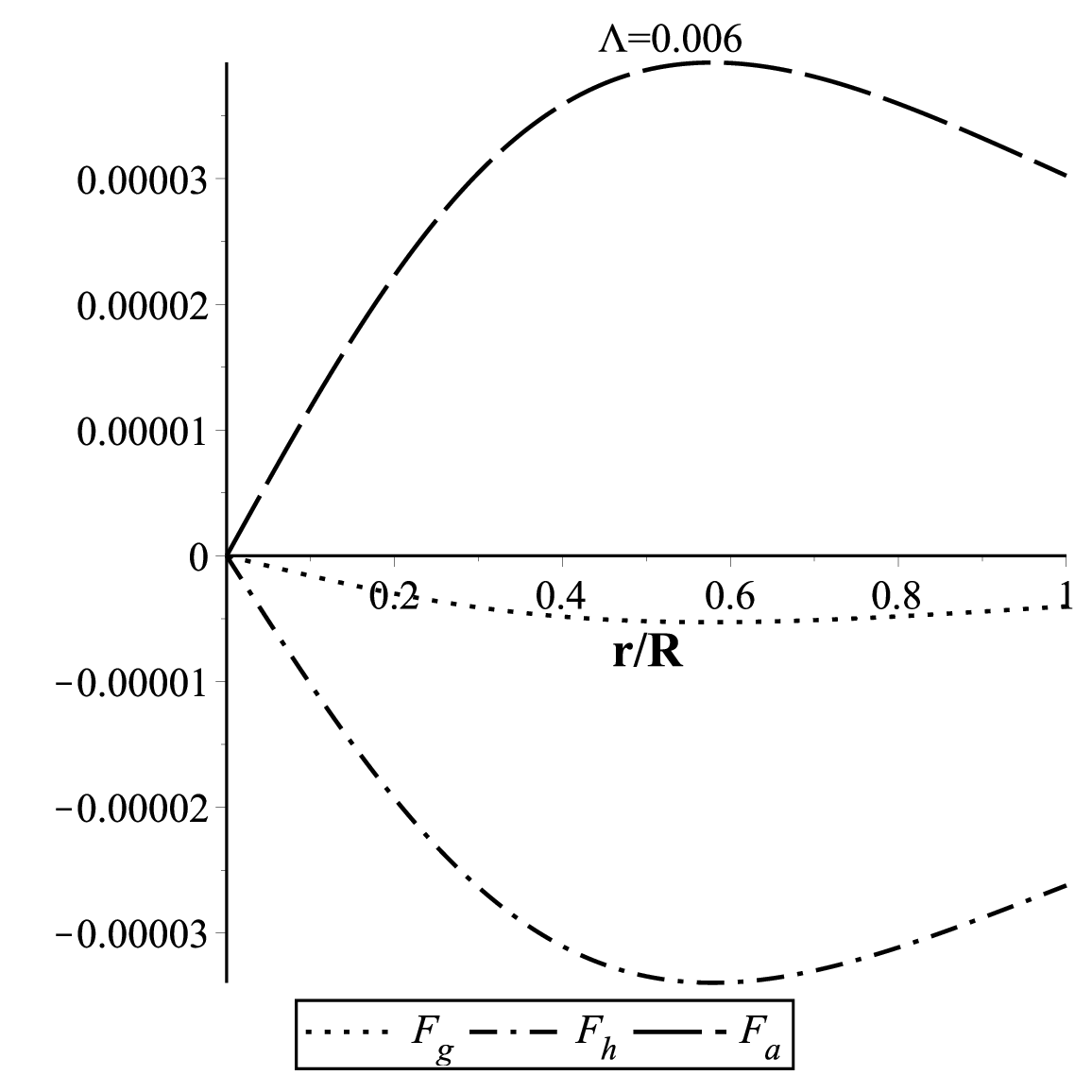}}
\caption{Behavior of different forces for $\Lambda=0.004$ (left panel), $\Lambda=0.005$ (middle panel) and $\Lambda=0.006$ (right panel) as a function of the fractional radial distance for the strange star candidate $LMC~X-4$ are shown}\label{FigTOV}
\end{figure}

Fig.~\ref{FigTOV} shows the counterbalancing pattern of different forces to attain equilibrium of the spherical configuration. Fig.~\ref{FigTOV} also features that in each cases the repulsive anisotropic force $F_a$ acts along the outward direction to counter balance the combined effect of the attractive forces viz., gravitational force $F_g$ and the hydrodynamic force $F_h$, which act along the inward direct.

\subsubsection{Herrera condition}
To verify stability of our anisotropic model we have plotted the radial ($v_{r}^2$) and transverse ($v_{t}^2$) sound
speeds in Fig. 7. It is observed that the inequalities $0\leq v_{r}^2 \leq 1$ and $0\leq v_{t}^2 \leq 1$ are satisfied everywhere within the stellar system and obeys the causality condition. Now the radial and transverse sound speeds are given by
\begin{eqnarray}
&\qquad v_{r}^2=\frac{dp_r}{d\rho}=\alpha,\\
&\qquad v_{t}^2=\frac{dp_t}{d\rho}  =\frac{-B^2 (1 + \alpha) + (1 + \alpha) a^3 r^2 +(1 + \alpha) b^3 r^8 +V_{t1}(r) +V_{t2} (r)+V_{t3} (r)}{2[a^2 (2 + B r^2) + a (B + 6 b r^2 + 3 b B r^4) -2 b (1 - 3 b r^4) + 2 b B r^2 (1 + b r^4)]},
\end{eqnarray}

where
\begin{eqnarray*}
&\qquad V_{t1}(r)= b [1 - 3 \alpha + 4 B r^2 (1+2\alpha)]+b^2 r^4 [6(\alpha-1)- 4 B r^2 + B^2 r^4 (1+\alpha)],\\
&\qquad V_{t2}(r)= - a^2 [1 -3 \alpha + B r^2 (1- \alpha)- 3 b r^4 (1+\alpha)]+a [B^2r^2 (1 +\alpha)  (b r^4-1)],\\
&\qquad V_{t3}(r)=a\,[ B (1 + 3 \alpha) - 3 b B r^4 (1- \alpha)] - a b r^2 [5 - 7 \alpha- 3 b r^4 (1+\alpha)].\\
\end{eqnarray*}

\begin{figure}[!htp]
\centering
\includegraphics[width=5cm]{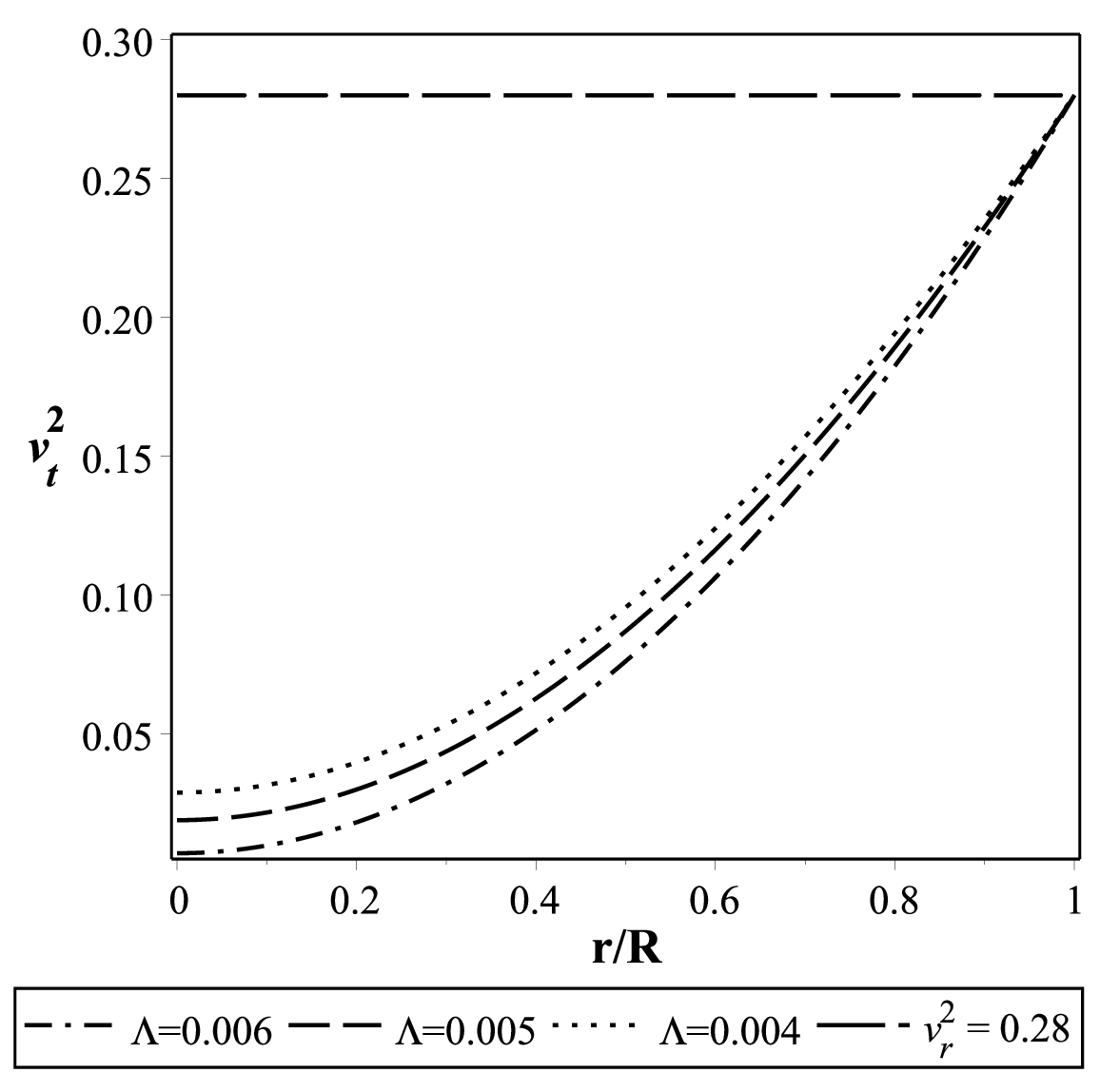}
\includegraphics[width=5cm]{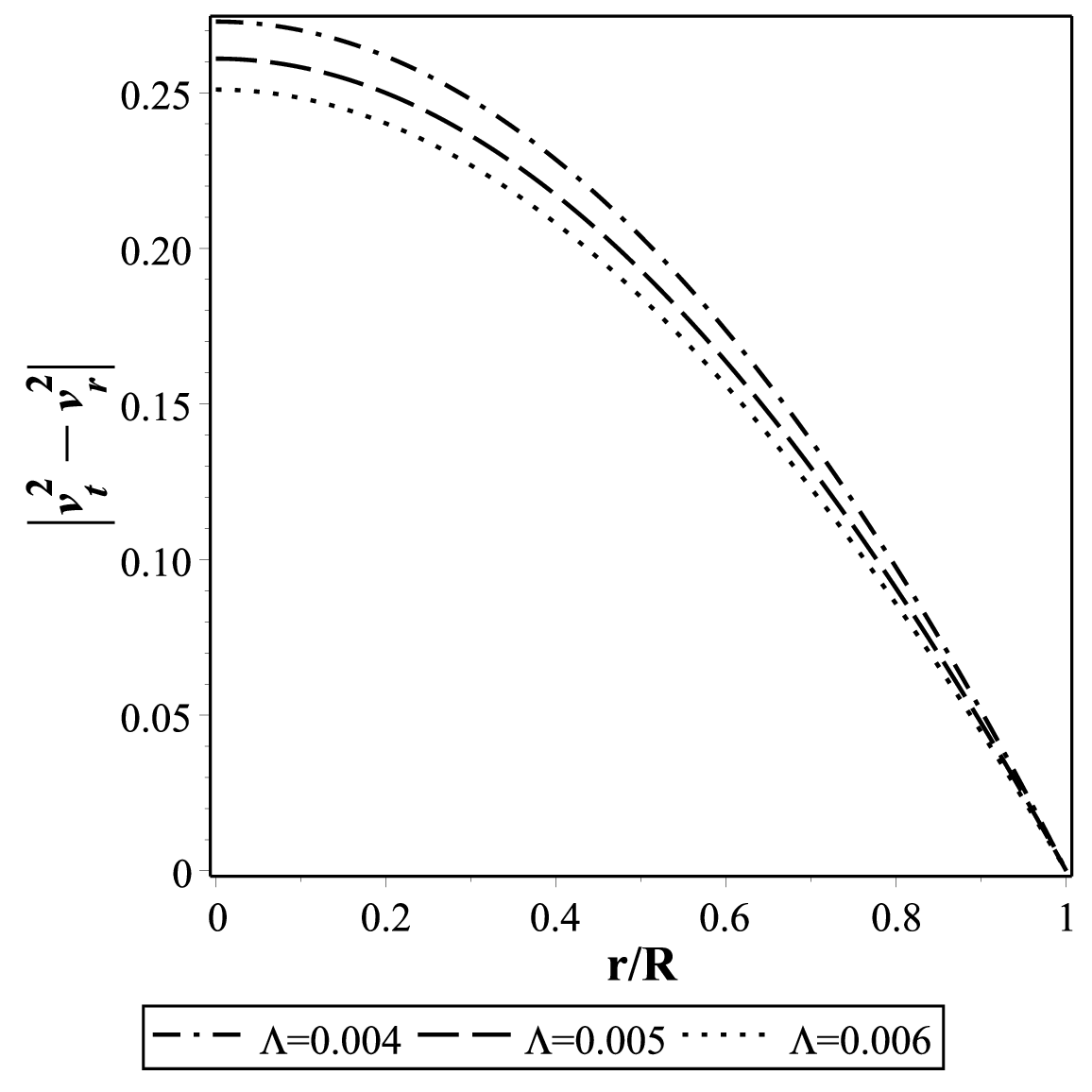}
\caption{Behavior of the square of the tangential sound speed $v_{t}^2$ (left panel) and difference of the square of the sound speeds  $\mid v_{t}^2 - v_{r}^2 \mid $ (right panel) as a function of the fractional radial distance for the strange star candidate $LMC~X-4$ are shown}\label{Figvel}
\end{figure}

On the other hand, following the cracking concept~\cite{Herrera1992,Abreu2007} we checked stability of the local anisotropic configuration by finding out the potentially stable region where the radial speed of sound is greater than the transverse speed of sound, i.e. $v^2_{t}-v^2_{r}\leq 1$. From Fig.~\ref{Figvel} one can note that $0\leq v_{r}^2 \leq 1$ and $0\leq v_{t}^2 \leq 1$ so that $\mid v_{t}^2 - v_{r}^2 \mid \leq 1 $. Thus the present compact stellar model provides a stable configuration of the spherical system by satisfying both the causality condition and the Herrera cracking concept.

\subsubsection{Adiabatic index}
To check the dynamical stability of the stellar system against an infinitesimal radial adiabatic perturbation we follow the pioneering work by Chandrasekhar~\cite{Chandrasekhar1964} where he predicted that for relativistic stellar system to be stable the adiabatic index $\Gamma$ should be greater than $4/3$. Note that later on many researchers~\cite{Hillebrandt1976,Horvat2010,Doneva2012,Silva2015} have successfully tested the prediction by Chandrasekhar for both the isotropic and anisotropic stellar objects. The adiabatic index is defined as
\begin{eqnarray*}
{\Gamma}=\left(\frac{{p_{r}}+{{\rho}}}{{p_{r}}}\right)\frac{d{p_{r}}}{d{{\rho}}}.
\end{eqnarray*}

\begin{figure}[!htp]
\centering
\includegraphics[width=6cm]{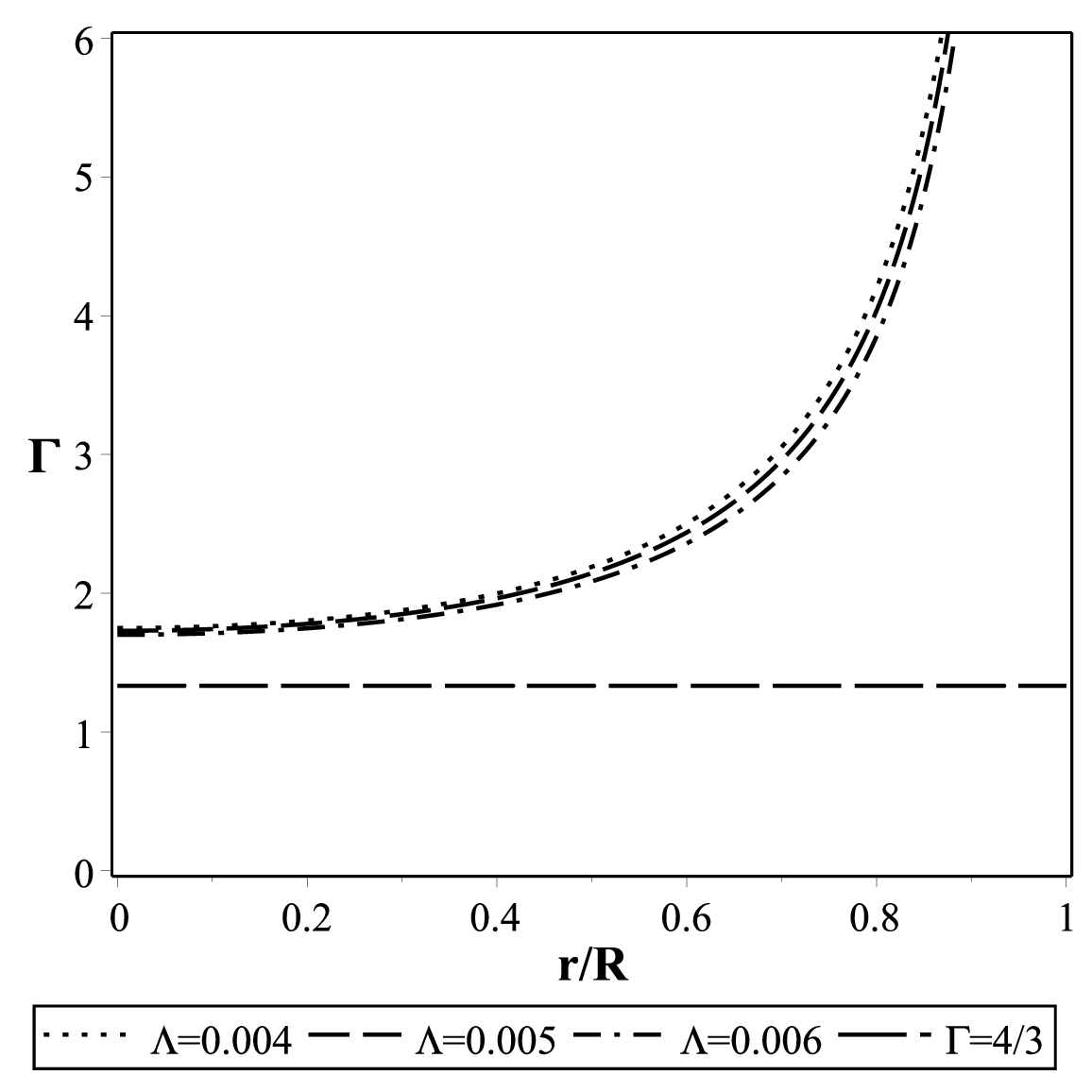}
\caption{Behavior of the adiabatic index $\Gamma$ as a function of the fractional radial distance for 
the strange star candidate $LMC~X-4$ is shown}\label{Figadia}
\end{figure}

In Fig.~\ref{Figadia} we have successfully shown the behaviour of the adiabatic index $\Gamma$ which features that in the present case $\Gamma$ is well above the critical limit $4/3$ and confirms stability of the system against an infinitesimal radial adiabatic perturbation.

\subsection{Compactfication factor and surface redshift}
The mass function of the present spherical stellar system is given as
\begin{eqnarray}
&\qquad\hspace{-9cm} m\left(r\right)=4\pi{\int^r_0{\rho r^2dr}} \nonumber\\
 &\qquad\hspace{-0.30cm} =8\,b \Big[ -\sqrt {\delta_{{1}}}\sqrt {2} \left( \chi_{{1}}-\chi_{{2}
} \right) {\rm arctanh} \left({\frac {\sqrt {b}r\sqrt {2}}{\sqrt {
\delta_{{2}}}}}\right)+\sqrt {\delta_{{2}}} \big\lbrace  \left( \chi_{{1}}+
\chi_{{2}} \right)\nonumber\\
 &\qquad\hspace{-0.30cm} \sqrt {2}\arctan \left( {\frac {\sqrt {b}r\sqrt {2}
}{\sqrt {\delta_{{1}}}}} \right)  +\frac{1}{3}b\,r\sqrt {{a}^{2}-4\,b}\sqrt {
\delta_{{1}}} \left( b\beta\,{r}^{6}+a\beta\,{r}^{4}+\beta\,{r}^{2}-3
 \right)  \big\rbrace  \Big] \nonumber\\
 &\qquad\hspace{-0.30cm} \Bigg/ \left[ {\delta_{{1}}}^{\frac{3}{2}}{\delta_{{2}}}^{\frac{3}{2}}\sqrt {{a}^{2}-4\,b} \left( 2\,b{r}^{2}+\delta_{{1}} \right)  \left( -2\,b{r}^{2}+\delta_{{2}}
 \right)  \left( \alpha+1 \right) 
 \right], \label{eq32}
\end{eqnarray}
where \\$\delta_{{1}}=\sqrt {{a}^{2}-4\,b}+a$,~$\delta_{{2}}=\sqrt {{a}^{2}-4\,b}-a$,~$\chi_{{1}}= \left\lbrace {r}^{4}{b}^{\frac{3}{2}}+\sqrt {b} \left( a{r}^{2}+1 \right)  \right\rbrace B\sqrt {{a}^{2}-4\,b}$,~$\chi_{{2}}= \left( Ba{r}^{4}-a{r}^{2}-1 \right) {b}^{\frac{3}{2}}-{r}^{4}{b}^{\frac{5}{2}}+Ba\sqrt {b} \left( a{r}^{2}+1 \right)$.

\begin{figure}[h]
\centering
\subfloat{\includegraphics[width=5cm]{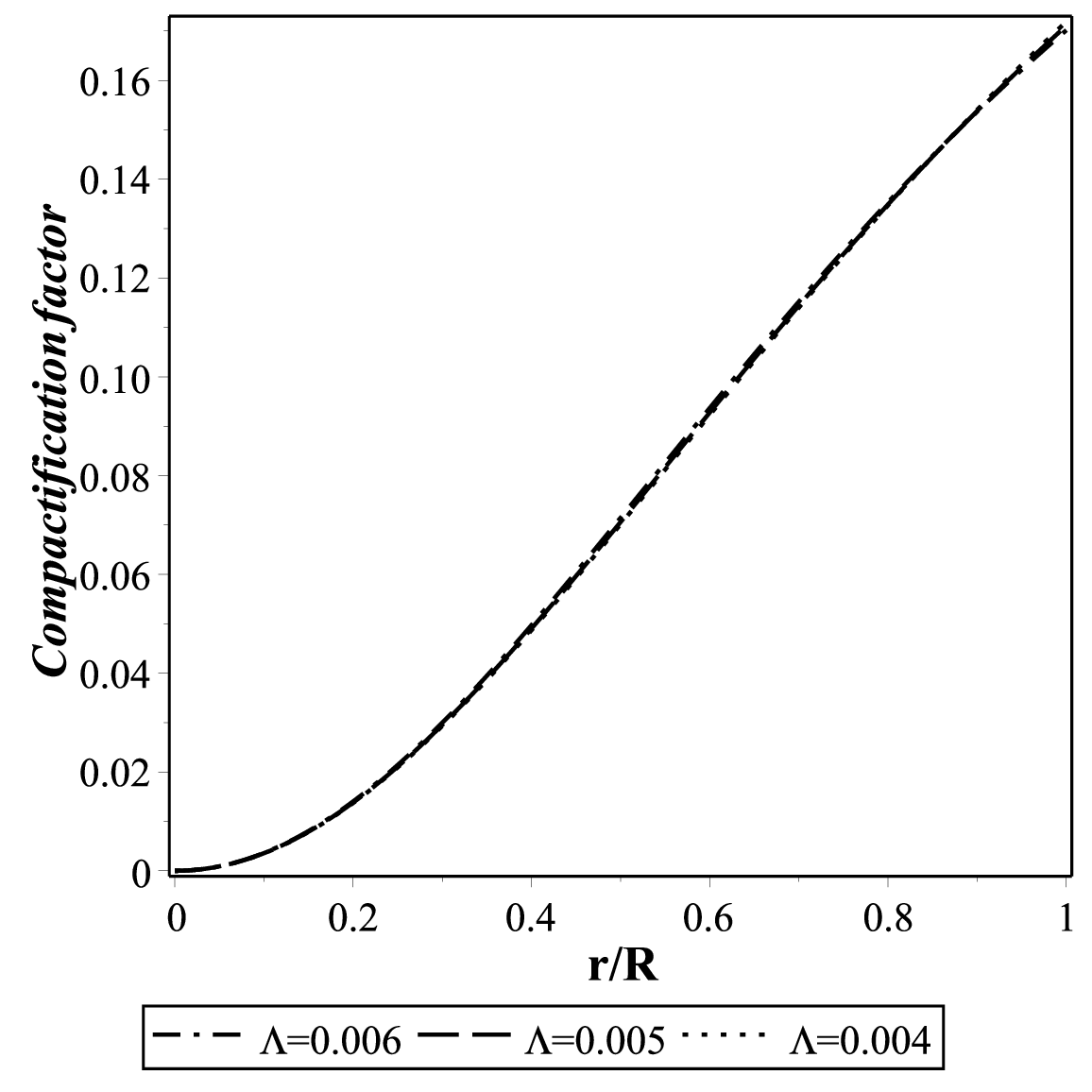}}
\
\subfloat{\includegraphics[width=5cm]{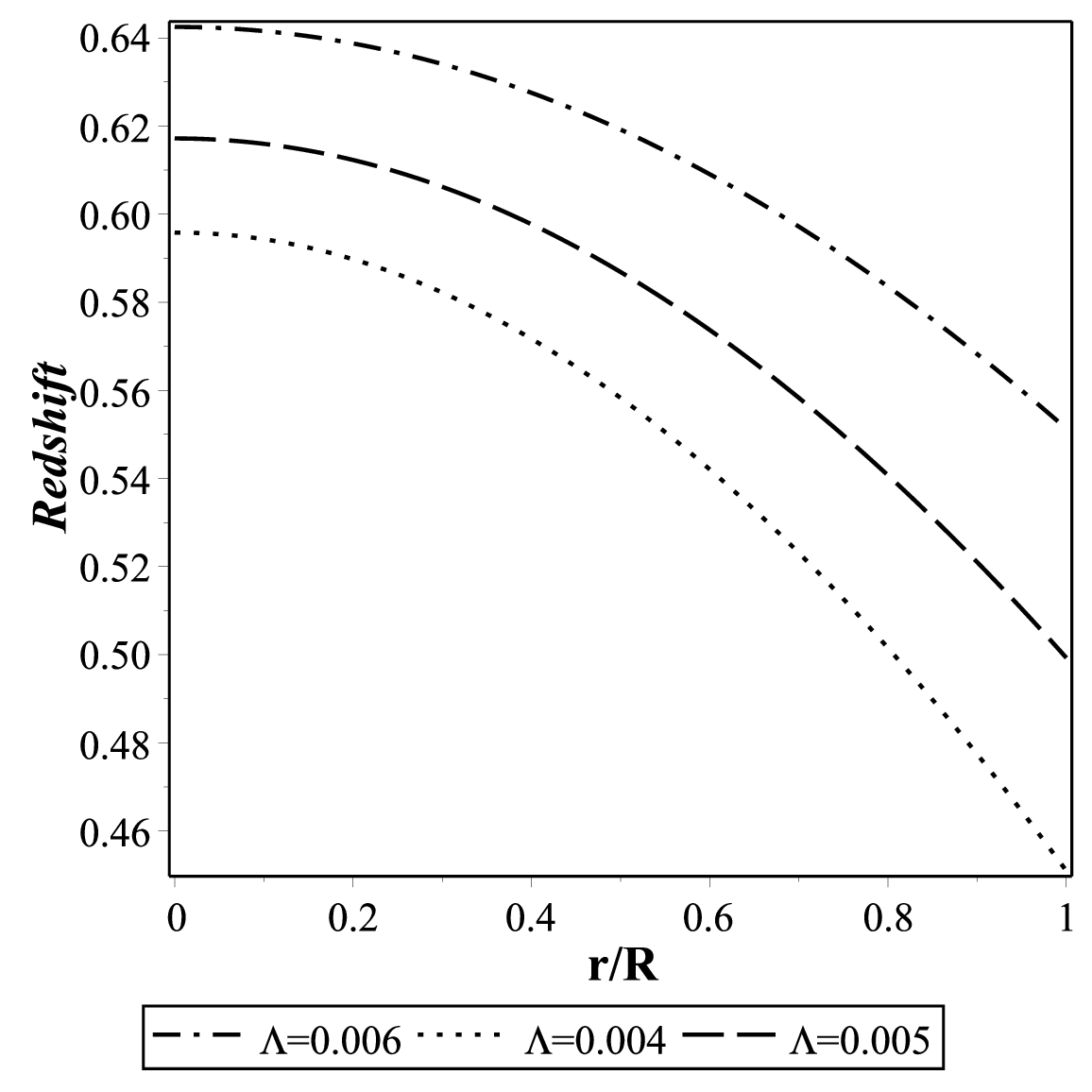}}
\caption{Behavior of the compactification factor (left panel), whereas behavior of the redshift (right panel) as a function of the fractional radial distance for the strange star candidate $LMC~X-4$ are shown }\label{FigRedshift}
\end{figure}

In terms of the mass function, let us now define the compactification factor as follows 
\begin{eqnarray}
u=\frac{m\left(r\right)}{r}.
\end{eqnarray}

Hence the surface redshift is given by 
\begin{eqnarray}
&\qquad\hspace{-3cm} Z_s={\rm e}^{-\frac{\nu\left(R\right)}{2}}-1=\left[1-2u\left(R\right)\right]^{-\frac{1}{2}} - 1.
&
\end{eqnarray}

We have plotted variation of the compactification factor and redshift function in the left and right panel of Fig.~\ref{FigRedshift}.

\section{Discussion and Conclusion}\label{sec7}
In the present study, we have considered an anisotropic fluid sphere in the framework of Einstein's general theory of relativity. The spherical spacetime is assumed to be of the Tolman-Kochowicz type and solutions of the Einstein field equations are found by using this metric. One can observe that all the parameters involved in the solutions set are viable within the specified physical conditions. Further, following the work of Stergioulas~\cite{Stergioulas2003} we have analyzed the present system based on the phenomenological MIT bag model EOS which is suitable for the massive strange quarks. To derive the unknown constants and radius of the strange star candidates in the present study we have used the observed values of the mass of the strange stars candidates as presented in Table~\ref{Table 1} and also assumed parametric values of $\mathcal{B}$ and $\Lambda$ given as $\mathcal{B}=50, 60$ and $70~MeV/{fm}^3$, and $\Lambda=0.004, 0.005$ and $0.006$. For further analysis of the achieved solutions, we have considered $LMC~X-4$ as the representative of the strange star candidates.


\begin{table*}
  \centering
    \caption{Numerical values of physical parameters for $LMC~X-4$ for the different values of $\Lambda$ and $\mathcal{B}=60 MeV/{fm}^3$ } \label{Table 2}
        \scalebox{0.9}{
\begin{tabular}{ ccccccccccccccccccccccccccc}
\hhline{=========}
Values  & Predicted  & $\rho_c$ & $p_c$  & $2M/R$ & $Z_s$ \\ 
of $\Lambda$ & Radius (km) & $\left( gm/{cm}^3\right)$ & $\left(dyne/{cm}^2\right)$ & &\\
\hline
0.004 & $9.275 \pm 0.116$ & $1.340\times {10}^{15}$ & $2.295\times {10}^{35}$ & $0.410$ & 0.302 \\ 

0.005 & $9.170 \pm 0.098$ & $1.380\times {10}^{15}$ & $2.396\times {10}^{35}$ & $0.415$ & 0.307 \\ 

0.006 & $9.033 \pm 0.087$ & $1.444\times {10}^{15}$ & $2.558\times {10}^{35}$ & $0.421$ & 0.314 \\ 

\hhline{=========} 
\end{tabular}  }
  \end{table*}



\begin{table*}
  \centering
    \caption{Numerical values of physical parameters for $LMC~X-4$ for the different values of $\mathcal{B}$ and $\Lambda=0.005$ } \label{Table 3}
        \scalebox{0.9}{
\begin{tabular}{ ccccccccccccccccccccccccccc}
\hhline{=========}
Values  & Predicted  & $\rho_c$ & $p_c$  & $2M/R$ & $Z_s$ \\ 
of $\mathcal{B}$ & Radius (km) & $\left( gm/{cm}^3\right)$ & $\left(dyne/{cm}^2\right)$ & &\\
\hline
50 & $9.659 \pm 0.105$ & $1.161\times {10}^{15}$ & $2.025\times {10}^{35}$ & $0.39$ & $0.28$ \\ 

60 & $9.170 \pm 0.098$ & $1.380\times {10}^{15}$ & $2.396\times {10}^{35}$ & $0.42$ & $0.31$ \\ 

70 & $8.757 \pm 0.093$ & $1.613\times {10}^{15}$ & $2.804\times {10}^{35}$ & $0.44$ & 0.34 \\ 

\hhline{=========} 
\end{tabular}  }
  \end{table*}


However, we would like to highlight some of the salient features of the present models which are as follows:

{\bf 1. Density and pressure:} We have shown the profile of density in Fig.~1, which features that for $\mathcal{B}=60~MeV/{fm}^3$ and the chosen values of $\Lambda$, the nuclear density inside the compact stellar model is much higher than the normal nuclear density ${\rho}_{normal}=2.3\times {10}^{14}~gm/{cm}^3$. Hence, it confirms that the stellar system corresponds to the ultra-dense strange stars~\cite{Ruderman1972,Glendenning1997,Herzog2011}. In Fig.~2 we have shown the variation of $p_r$ and $p_t$ in the left and right panel, respectively. On the other hand, we have shown variation of the anisotropic stress in Fig.~4, which indicates that the anisotropy for our system is minimum at the centre and maximum at the surface as predicted by Deb et al.~\cite{Deb2017}. Interestingly, Fig.~4 also features that as the values of $\Lambda$ increase anisotropic stress of the systems increase consequently. We have presented the variation of the cosmological constant which is a scalar variable dependent on the spatial coordinate in Fig.~3. The variations of the metric potentials, i.e., $e^{\nu}$ and $e^{\lambda}$ are depicted in the left and right panel of Fig.~5, respectively. Further, Figs.~1, 2 and 5 confirm that our system is free from all sorts of singularities viz., physical as well as geometrical singularities.

{\bf 2. Energy conditions:} We find that for both the chosen cases the energy conditions, viz., NEC, WEC, SEC and DEC are consistent with our system. We have shown variation of the energy conditions with respect to the radial coordinate in Fig.~6, which confirms physical acceptability  of the obtained solutions.

{\bf 3. Equilibrium of forces:} In Fig.~7 we observe that for all the chosen values of $\Lambda$ the equilibrium of the forces is achieved. In every cases the anisotropic force $F_a$ which acts along the outward direction (repulsive nature) counterbalances the combined effect of the gravitational force $F_g$ and the hydrodynamic force $F_h$ which act along the inward direction (attractive nature).

{\bf 4. Stability:} To discuss stability of the system we have examined both the causality condition and the Herrera cracking condition. Fig.~8 exhibits that for all the cases the inequalities $0\leq v_{r}^2 \leq 1$, $0\leq v_{t}^2 \leq 1$ and $\mid v_{t}^2 - v_{r}^2 \mid \leq 1 $ are valid simultaneously and hence confirms stability of the stellar system in terms of the sound velocity of the system. Again, Fig.~9 reveals that for our system the adiabatic index $\Gamma$ is less than the critical limit $4/3$ which confirms that our system is completely stable against the infinitesimal radial adiabatic perturbation.

{\bf 5. Compactification factor and redshift:} We have featured variation of the compactification factor and redshift function in the left and right panel in Fig.~10, respectively. For the chosen values of $\mathcal{B}$ and $\Lambda$ we have derived the factor $2M/R$ and surface redshift $Z_s$ for different strange star candidates and presented their values in Tables~\ref{Table 1}. We find from these Tables~\ref{Table 2} and \ref{Table 3} that the values of $2M/R$ and $Z_s$ increase gradually with the increasing values of $\Lambda$ and $\mathcal{B}$. For all the cases as $2M/R<8/9$, so our system is consistent with the Buchdahl condition~\cite{Buchdahl1959}. Present investigation reveals that the surface redshift of the different stellar candidates are within the range $0.58-0.41$ and such high redshift values indicate that the stellar candidates under study are actually stange stars~\cite{Rahaman2014}. The obtained values of $Z_s$ due to different strange star candidates are well within the provided limits given as $0<Z_s \leq 1$~\cite{Kalam2012,Hossein2012,Rahaman2012,Kalam2013,Bhar2015} and $Z_s \leq 2$~\cite{Buchdahl1959,Straumann1984,Bohmer2006} which provides physical consistency of the proposed stellar system.

{\bf 6. Comparative discussion:} With the motivation to analysis the effects of the variable cosmological constant $\Lambda(r)$ and bag constant $\mathcal{B}$ on the anisotropic stellar system, we have presented Tables~\ref{Table 2} and \ref{Table 3}. For the fixed chosen value of $\mathcal{B}$ as $\mathcal{B}=60~MeV/{fm}^3$ we find from Table~\ref{Table 2} that as the values of $\Lambda$ increases the different physical parameters viz., $\rho_c$, $p_c$, $2M/R$ and $Z_s$ increase gradually, whereas the radii of the stellar candidates decrease consequently. This observation leads to the conclusion that as $\Lambda$ increases the stellar systems become smaller in size turning the strange star candidates into a more compact stellar object. The same effects can be observed for the increasing values of $\mathcal{B}$ as presented in Table~\ref{Table 3}. Hence, with the increasing values of $\Lambda$ and $\mathcal{B}$ the systems become gradually diminutive along with the rising values of density and surface redshift, which are the perfect situations to describe the strange star candidates.

The pioneering detection of gravitational waves (GW170817)~\cite{Abbott2017} from the coalescence of a neutron star binary system, has given a realistic opportunity to study the properties of the matter inside the stars in such extreme condition which cannot be recreated in the terrestrial laboratories. In their recent study, Abbott et al.~\cite{Abbott2018} have presented the radii of the two neutron stars. The authors have shown that based on only the LIGO and Virgo data the two neutron star radii are $R_1=10.8^{+2.0}_{-1.7}$ km and $R_2=10.7^{+2.1}_{-1.5}$ km for the heavier and lighter stars, respectively at the $90\%$ confidence level. However, for the neutron stars having masses more than $1.97~M_{\odot}$ which includes the electromagnetic observations Abbott et al.~\cite{Abbott2018} predicted that radii of the two neutron stars are exactly the same, i.e. $R_1=11.9^{+1.4}_{-1.4}$ km and $R_2=11.9^{+1.4}_{-1.4}$ km at the $90\%$ confidence level. They also predicted that at twice the nuclear saturation density the pressure should be $3.5^{+2.7}_{-1.7}\times {10}^{34} dyne/{cm}^2 $. In the present study, our model reveals that for a hypothetical strange quark star having mass $2.2~M_{\odot}$ (say) the radius would be $9.981$ km and central pressure would be $4.836 \times {10}^{35} dyne/{cm}^2$, where the value of $B$ is chosen as $60~MeV/{fm}^3$. It is to note that our model predicts the stars having more dense and smaller in size compared to the neutron stars, however, this situation can easily be well accomodated within the range of data as predicted by Abbott et al.~\cite{Abbott2018}.

For the last few years, the scalar-tensor theories of gravity become popular as an extended theory of General Relativity (GR) to address several puzzles, such as the dark energy problem in the realm of Cosmology and the dark matter problem in the realm of astrophysics. In this line, $f(R)$ gravity theory has appeared as one of the promising alternative theories to address those unsolved issues which are studied in detail in the following literature~\cite{Dombriz2006,Dombriz2008,Dombriz2011,Dombriz2012,Dombriz2012a,Dombriz2012b,Dombriz2013,Dombriz2013a,Dombriz2016,Dombriz2016a}. 
Now, our present study deals with the cosmological constant $\Lambda$ which is assumed to be a scalar variable dependent on the spatial coordinate $r$. Interestingly, we find that the violation of standard GR predictions has much in common with the results achieved by Resco et al.~\cite{Dombriz2016a} in their work where they have explored the main features and also studied the existence of neutron stars in the framework of $f(R)$ paradigmatic models. Hence, the achieved results from our model have actually encouraged the extended gravity theories as a competent and alternative route to the cosmological constant or dark energy. However, following Dombriz and Dobado~\cite{Dombriz2006} we are looking forward to further detailed study on strange stars based on the assumed Tolman-Kuchowicz metric in the framework of extended gravity theory without considering the cosmological constant.

In this context, it is important to mention that though in the present study under the framework of GR we have used the standard asymptotic (or stellar) mass definition as in Eq. (5) [$m \left( r \right) =4\pi\int_{0}^{r}\!\rho \left( r \right) {r}^{2}{dr}$] but for modified $f(R)$ gravity theory this standard definition of mass should be avoided. Astashenok et al.~\cite{Astashenok2017} in their study have shown that in the case of modified $f(R)$ gravity theory as at the surface of the stars the scalar curvature $R$ does not vanish hence it invites not only the existence of exterior non-Schwarzschild solutions for realistic equations of state but also a new definition for the gravitational mass of the stellar system. As a result, the exterior spacetime region (gravitational sphere) of the stars also contribute to the total gravitational mass which can be perceived by a distant observer.

As a final comment, in this article employing the Tolman-Kuchowicz~\cite{Tolman1939,Kuchowicz1968} metric and using parametric values of $\Lambda(r)$ and $\mathcal{B}$ we have successfully presented singularity free and completely stable stellar system which is suitable to describe the strange star candidates exhibiting anisotropic nature.

\section*{Acknowledgments}
SR is thankful to the Inter-University Centre for Astronomy and Astrophysics (IUCAA), Pune, India and The Institute of Mathematical Sciences, Chennai, India for providing all types of working facility and hospitality under the Associateship scheme. A part of this work was completed while D.D. was visiting IUCAA and the author gratefully acknowledges the warm hospitality and facilities in the library there. We all are thankful to the anonymous referee for the pertinent comments which has helped us to upgrade the manuscript substantially.

\end{document}